\documentclass[aps,twocolumn,showpacs,superscriptaddress]{revtex4}
\usepackage{epsfig}
\usepackage{graphicx}
\usepackage{amsfonts}
\usepackage[figuresright]{rotating} 
\usepackage{amssymb}
\usepackage{amsmath}
\usepackage{psfrag}

\def\avg#1{\langle#1\rangle}

\def\be{\begin{equation}} \def\ee{\end{equation}}
\def\bea{\begin{eqnarray}} \def\eea{\end{eqnarray}}

\def\nn{\nonumber}
\def\pp{\parallel}

\begin{document}
\title{The $p_{x,y}$-orbital counterpart of graphene: 
cold atoms in the honeycomb optical lattice}

\author{Congjun Wu}
\affiliation{Department of Physics, University of California, San Diego,
CA 92093}
\author{S. Das Sarma}
\affiliation{Condensed Matter Theory Center, 
Department of Physics, University of Maryland, College Park, MD 20742} 

\begin{abstract}
We study the ground state properties of the interacting spinless fermions
in the $p_{x,y}$-orbital bands in the two dimensional honeycomb optical
lattice, which exhibit different novel features from those in
the $p_z$-orbital system of graphene.
In addition to two dispersive bands with Dirac cones,
the tight-binding band structure exhibits another two completely 
flat bands over the entire  Brillouin  zone.
With the realistic sinusoidal optical potential, the flat bands acquire a
finite but much smaller band width compared to the dispersive bands.
The band flatness dramatically enhanced interaction effects
giving rise to various charge and bond ordered states at commensurate 
fillings of $n=\frac{i}{6}~ (i=1\sim 6)$.
At $n=\frac{1}{6}$, the many-body ground states can be exactly 
solved as the close packed hexagon states which can be stabilized
even in the weakly interacting regime.
The dimerization of bonding strength occurs at both $n=\frac{1}{2}$ and 
$\frac{5}{6}$, and the latter case is accompanied with the charge density 
wave of holes.
The trimerization of bonding strength and charge inhomogeneity appear at
$n=\frac{1}{3},\frac{2}{3}$.
These crystalline orders exhibit themselves in the 
noise correlations of the time of flight spectra.
\end{abstract}
\maketitle

\section{Introduction}
There has been tremendous progress during the past decade in the
cold atom physics.
In the early days, Bose-Einstein condensation was first realized
in magnetic traps by using dilute alkali atoms \cite{anderson1995,davis1995},
where interaction effects are weak.
Later on, important achievements have been made to realize strongly
correlated systems by using optical lattices.
The major advantage of optical lattices is the excellent 
controllability of interaction strength.
For example, the superfluid to Mott insulator transition of
bosons  has been experimentally observed \cite{greiner2002}. 
Recently, cold atom physics in optical lattices is merging with
condensed matter physics, which  provides a wonderful opportunity
to explore new states of matter.

An important aspect of strongly correlated systems 
is orbital physics, which studies an additional degree of
freedom independent of charge and spin.
In many transition metal oxides, the $d$-orbitals are partially filled,
which enables the orbital degree of freedom active.
Orbital physics is characterized by orbital degeneracy and 
spatial anisotropy of orbital orientation.
The interplay between orbital, spin and charge degrees of freedom
gives rise to many interesting phenomena such as metal-insulator transitions, 
superconductivity, and colossal magneto-resistance 
\cite{Imada1998,tokura2000,khaliullin2005}.

Orbital degrees of freedom also exist in optical lattices.
Although most of current research of cold fermions and bosons
focuses on the lowest $s$-orbital bands, large progress has been made
in high orbital bands.
An important advantage of optical lattices is the rigidity of lattices.
They are free of the Jahn-Teller type lattice distortion
which often occurs in transition metal oxides and quench the orbital
degrees of freedom.
Orbital physics in optical lattices  exhibits new 
features which are not usually realized in solid state systems. 
Recently, the properties of bosons in the first excited $p$-orbital
bands have been attracting a great deal of
attention \cite{scarola2005, isacsson2005, liu2006, kuklov2006, wu2006,
xu2007, xu2007a, alon2005}.
Scarola {\it et al.} proposed to realize the supersolid state
by using bosons in the high orbitals to generate the next-nearest 
neighbor interaction \cite{scarola2005}.
Isacsson {\it et al.} investigated the sub-extensive $Z_2$ symmetry of
the $p$-orbital bosons in the square lattice and its consequential nematic
superfluidity \cite{isacsson2005}.
Liu and Wu \cite{liu2006}, and  Kuklov \cite{kuklov2006} studied the
antiferromagnetic ordering of the on-site orbital angular momentum
moment.
It was also proposed in Ref. \cite{liu2006} to enhance the 
life-time of $p$-orbital 
bosons by using a Bose-Fermi mixture to reduce the available phase space
of decay process of bosons.
Wu {\it et al.} \cite{wu2006} further investigated the superfluid and Mott
insulating states of $p$-orbital bosons in the frustrated triangular lattice,
and found a novel  stripe phase of orbital angular 
momentums.
Xu {\it et al.} studied a model of bond algebraic liquid
phase \cite{xu2007} and
phase transitions in anisotropic $xy$-models \cite{xu2007a} 
in the context in the $p$-orbital boson systems.

On the experimental side, the progress of orbital physics with
cold atoms has also been truly exciting, which
opens up the new opportunity to study orbital physics.
Browaeys {\it et al.}
\cite{browaeys2005} and K\"ohl {\it et al.} \cite{kohl2005} have demonstrated
the population of high orbital bands with both bosons and fermions.
Furthermore, Sebby-strabley {\it et al.} \cite{sebby-strabley2006}
have  successfully pumped bosons into the excited bands in the
double-well lattice.
More recently, an exciting progress has been made by Mueller {\it et al.} 
\cite{mueller2007} to realize the meta-stable
$p$-orbital boson systems by using the stimulated Raman 
transition to pump bosons to high orbital bands.
The spatially anisotropic phase coherence pattern has been observed
in the time of flight experiments.
This opens up a new experimental direction to investigate novel 
condensate of bosons in the excited $p$-bands.

On the other hand, fermions in the $p$-orbital bands also possess
interesting behaviors \cite{wu2007,wu2008,wuzhai2007,zhao2008}.
Recently, Wu {\it et al.} \cite{wu2007} studied the flat band structure in 
the $p_{x,y}$-orbital physics in the honeycomb lattice.
Compared to the $p_z$-orbital system of graphene, which has been
attracting tremendous attention since the discovery of quantum 
Hall effect therein  \cite{novoselov2005,zhang2005,neto2007},
the $p_{x,y}$-orbital honeycomb systems exhibit new and even richer physics.
In graphene, the active bands near the Fermi energy are ``$\pi$''-type,  
composed  of the $p_z$-orbital directly normal to the graphene plane,
thus graphene is not a good system to investigate orbital physics.
In contrast, it is the other two $p$-orbitals ($p_{x,y}$) that lie in-plane
and exhibit both orbital degeneracy and spatial anisotropy,
giving rise to the interesting flat band physics \cite{wu2007}.
In solid state systems of graphene and MgB$_2$,
$p_{x,y}$-orbitals hybridize with the $s$-orbital, resulting
the $\sigma$-bonding (the $sp^2$ hybridization) band.
This $\sigma$-band  is fully filled and inert in graphene,
but is  partly filled and contributes to 
the two-band superconductivity in MgB$_2$ \cite{choi2002}.
Due to the large $s$-orbital component, 
the essential feature of orbital physics, orbital anisotropy, is not
prominent in these two systems.
In contrast, the $p_{x,y}$-orbital bands in optical lattices are well
separated from the $s$-band with negligible hybridization,
providing an unique opportunity to study the pure $p_{x,y}$-orbital
physics in the honeycomb lattice.
This research will provide us another perspective in the honeycomb lattice
and is complementary to the recent research focus on the single band
system of graphene. 
Other works of the $p$-orbital fermions include the investigation
of orbital exchange physics in the Mott-insulating states finding
various orbital ordering and frustration behavior
\cite{wu2008,zhao2008},
and the study of the possibility to enhance the antiferromagnetic 
ordering of fermions 
in the $p$-orbital of 3D cubic lattices \cite{wuzhai2007}.

Interaction effects in the $p_{x,y}$-orbital 
honeycomb optical lattices can be much stronger that those in 
the $p_z$-orbital graphene systems.
In real graphene the dimensionless coupling constant 
$r_s=e^2/(\epsilon\hbar v) $ has a maximum value of 2.3 
in vacuum (and $r_s<1$ for the current available graphene samples 
on SiO$_2$ or SiC substrates), taking $v=10^6$ cm/sec.
Thus graphene is very far from the $r_s=39$ regime needed for 
Wigner crystallization \cite{tanatar1989}.
Much of graphene interaction physics is  described by
perturbative weak-coupling renormalizations of the quasiparticle spectral
function, as shown both theoretically and experimentally 
\cite{dassarma2007,dassarma2007a,hwang2007}.
Furthermore, real graphene  physics is complicated by 
electron-phonon interactions \cite{tse2007}.
In contrast, in the $p_{x,y}$-orbital honeycomb lattices systems,
the flat band quenches the kinetic energy, and thus interaction physics is
non-perturbative and generic, leading to qualitatively new orbital physics
phenomena, e.g. Wigner-Mott physics, can show up easily
\cite{wu2007}.

This paper works as an expanded version of  
a previous publication of Ref. \cite{wu2007},
with new results and all the theoretical
details of the behavior of spinless fermions
in the $p_{x,y}$-orbital bands in the honeycomb lattice.
The current work is motivated by considerations of using the optical 
lattices to go beyond what can be achieved in solid state systems, i.e. 
obtain exotic strongly correlated orbital quantum phases which
have not yet been studied in condensed matter physics.
The paper is organized as follows.
In Sect. \ref{sect:band}, we analyze the band structures in both
the simplified tight-binding model and the realistic 
optical potential constructed from three co-planar laser beams 
\cite{grynberg1993}.
The band structures contain both Dirac cones in two dispersive bands,
and other two nearly flat bands over the entire Brillouin zone (BZ)
whose flatness becomes exact if the $\pi$-bonding is neglected.
Special attention is paid for the orbital configurations of
the localized Wannier-like eigen-functions in the flat bands,
and also at the Dirac points.
In Sect. \ref{sect:interaction}, the interacting Hamiltonian is introduced
and  methods of enhancing the Hubbard-like
on-site interaction  are proposed.
In Sect. \ref{sect:1/6}, the interaction effect in the partially-filled
flat band is discussed.
The situation is somewhat analogous to that in the fractional
quantum Hall effect  of electrons in the lowest Landau level.
When the flat band is partially-filled, the effects of interactions 
are entirely non-perturbative.  
We obtain the {\sl exact} many-body
plaquette Wigner crystal state at filling $\avg{n} =\frac{1}{6}$,
which is the close packed hexagon state and is stable even in 
the weak interaction regime.
In Sect. \ref{sect:comm}, we present various charge and bond
ordered states, including dimerized and trimerized states
at higher commensurate fillings in the strong interaction regime.
In Sect. \ref{sect:exp}, the
noise correlation in the time of flight experiments
are discussed.
Conclusions and outlook for the future research
are discussed in Sect. \ref{sect:conclusion}.

\section{$p_{x,y}$-orbital Band structure in the  honeycomb lattice}
\label{sect:band}
In this section, we will give a detailed analysis to the $p_{x,y}$-orbital
band structure in the 2D honeycomb lattice which is featured by
the interesting properties of both flat bands and Dirac cones.
We will first discuss the experimental construction of such a lattice
and then solve the band structure by using both the simplified
tight-binding model and the realistic sinusoidal optical potential.

\subsection{Construction of the optical honeycomb lattice}

\begin{figure}
\centering\epsfig{file=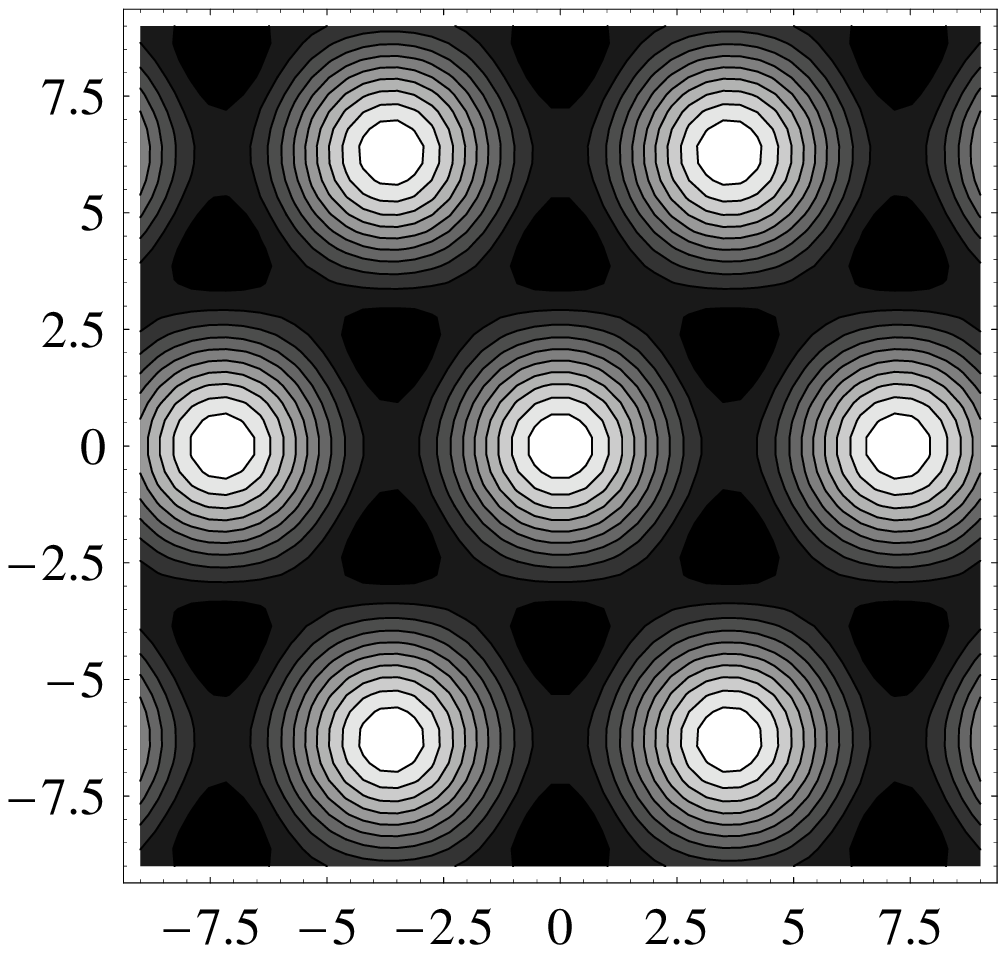,clip=1,width=0.45\linewidth,angle=0}
\ \ \
\centering\epsfig{file=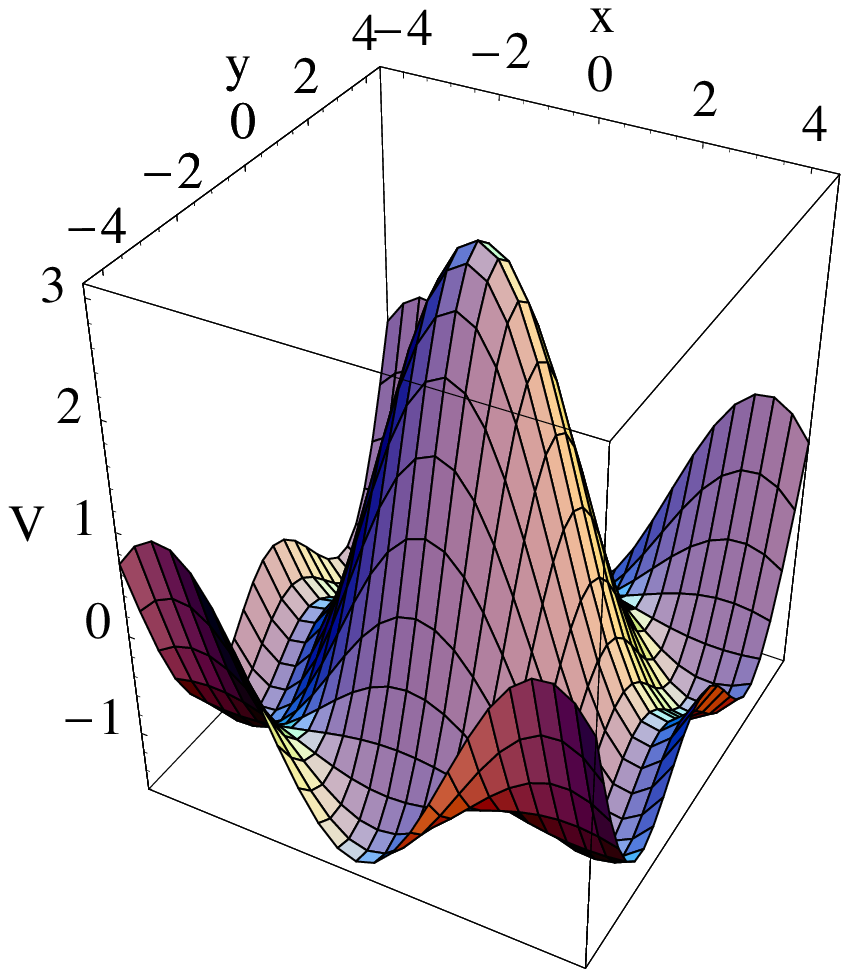,clip=1,width=0.40\linewidth,angle=0}
\caption{A) The contour plot of the optical potential of 
the 2D honeycomb lattice described by Eq. \ref{eq:potential}.
B) The optical potential distribution around the potential
maximum in one unit cell.
\label{fig:potential}
}
\end{figure}

The honeycomb optical lattice was realized experimentally 
by using three laser beams with co-planar propagating wavevectors
$\vec q_i (i=1 \sim 3)$ quite some time ago  \cite{grynberg1993}.
The magnitudes of these wavevectors are the same and their directions
form the angle of $120^\circ$ with each other.
Assuming the polarization of the electric fields of the three beams
are all along the $z$-direction, the optical potential distribution
can be expressed as
\bea
V(\vec r)= V_0 \sum_{i=1\sim 3} \cos (\vec p_i \cdot \vec r),
\label{eq:potential}
\eea
where $\vec p_1=\vec q_2 -\vec q_3$, $\vec p_2=\vec q_3 -\vec q_1$,
and $\vec p_3=\vec q_1 -\vec q_2$.
In the case of blue detuning, $V_0$ is positive and the potential
minima form a hexagonal lattice as depicted in Fig. \ref{fig:potential}.
In contrast, the red detuning laser beams generate a 2D triangular
lattice.
In both cases, the lattice is topologically stable against the phase
drift of the laser beams, which only causes a overall shift but not
the distortion of the lattice.
Fig. \ref{fig:potential} B depicts the potential distribution in one
unit cell of the honeycomb lattice, where a potential maximum 
locates in the center and six potential minima sit around.
Without loss of any generality, we take $\vec p_{1,2}=p(\pm \frac{\sqrt 3}{2}
\hat e_x +\frac{1}{2} \hat e_y)$ and $\vec p_3= - p~ \hat e_y$ where
$p=\frac{4\pi}{3a}$ and $a$ is the distance between the nearest neighbour
site in the honeycomb lattice.
We define the recoil energy in such a lattice
system as $E_r=\frac{\hbar^2 p^2}{2m}$ where $m$ is the mass of the  atom.

\subsection{The tight-binding model }
\begin{figure}
\psfrag{e}{$\hat e$}
\centering\epsfig{file=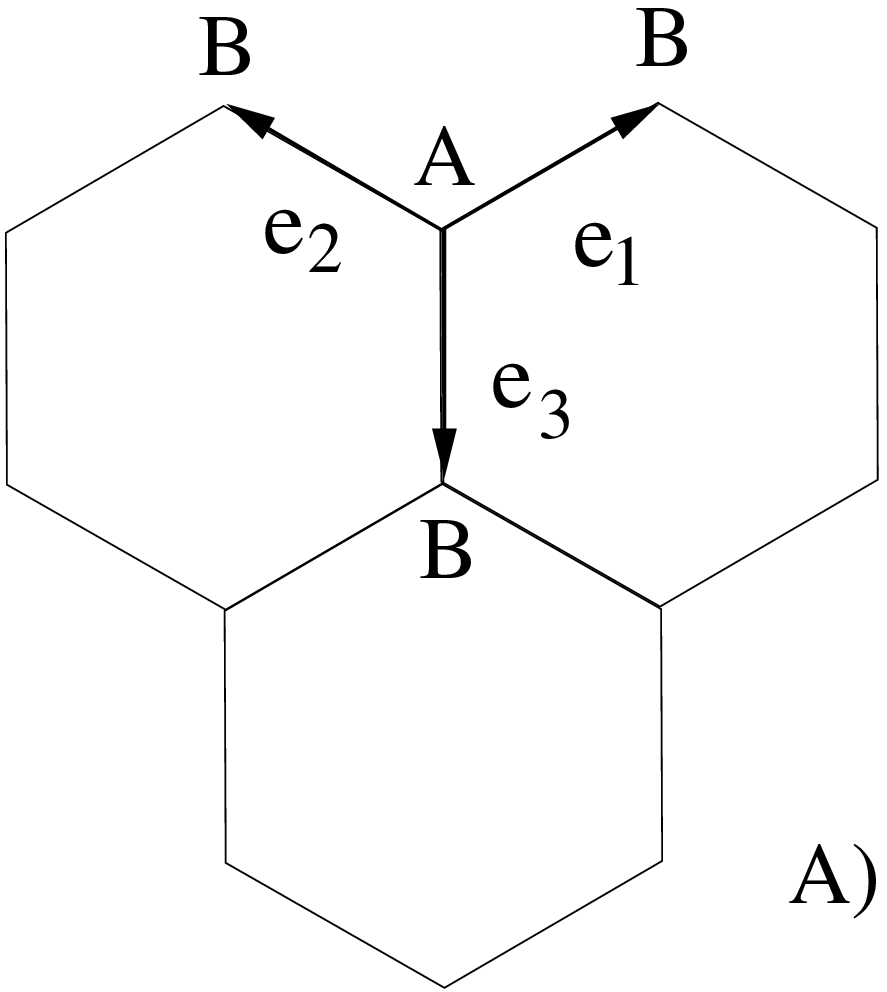,clip=1,width=0.35\linewidth,angle=0}
\ \ \, \ \ \,
\centering\epsfig{file=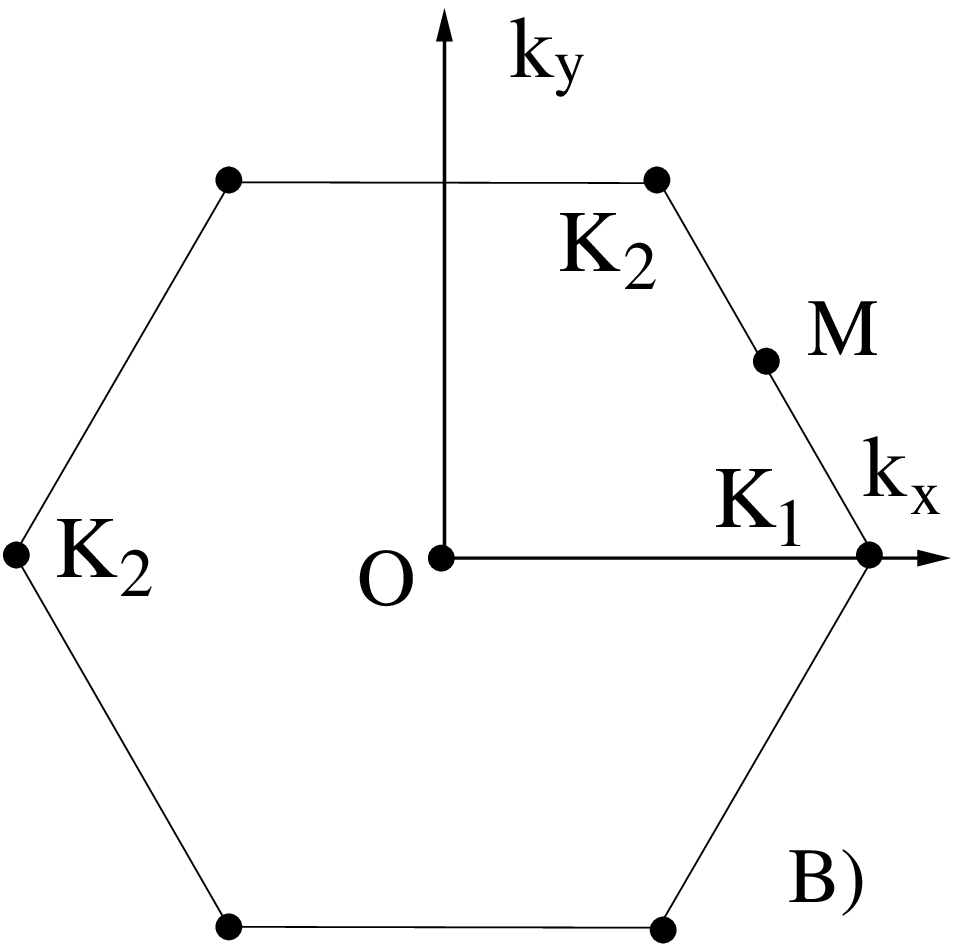,clip=1,width=0.4\linewidth,angle=0}
\caption{A) The two sublattice structure ($A$ and $B$) of the
honeycomb lattice.
B) The hexagon Brillouin zone with edge length $4\pi/(3\sqrt 3 a)$.
\label{fig:hnycmb_lat}
}
\end{figure}

\begin{figure}
\centering\epsfig{file=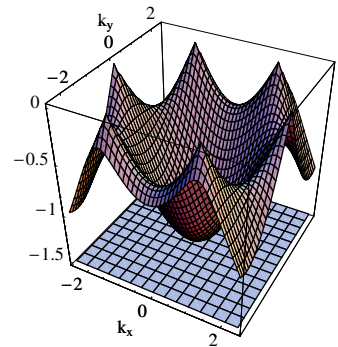,clip=1,width=0.5\linewidth,angle=0}
\caption{Dispersion of the two-lowest $p_{x,y}$-orbital bands
  $E_{1,2}$. 
  The band $E_1$ is completely flat, while $E_2$ exhibits Dirac points at
  $K_{1,2}=(\pm\frac{4\pi}{3\sqrt 3 a},0)$. The other two bands are
  symmetric with respect to $E=0$. From Wu  {\it et al} \cite{wu2007}.}
\label{fig:hnycmb_band}
\end{figure}

The optical potential around the center of each site is approximately
an anisotropic harmonic well.
We assume that the vibration frequencies along the $x$, $y$ 
and $z$-directions 
satisfy $\omega_z \gg \omega_x=\omega_y=\omega_{xy}$, and thus the 
energy of the $p_z$-orbital is much higher than that of 
the $p_{x,y}$-orbital bands.
When the lowest $s$-band is fully filled and thus inert, the active 
orbital bands will be of the $p_{x,y}$.
Due to the spatial orientation of the $p$-orbitals, the hopping processes 
in the $p$-orbitals can be classified into the $\sigma$ and $\pi$-type
bondings, respectively.
The former describes the hopping between $p$-orbitals on neighbouring
sites with the orientation along the bond direction, while the latter 
describes the hopping between $p$-orbitals perpendicular to the bond 
direction.
In other words, the $\sigma$-bonding is of the ``head to tail'' type, 
while the $\pi$-bonding is of the ``shoulder by shoulder'' type.
Typically, the amplitude of the $\pi$-bonding is much smaller than
that of the $\sigma$-bonding because of the strong orientational anisotropy.

The structure of the honeycomb lattice is depicted in Fig. 
\ref{fig:hnycmb_lat} A.
Each unit cell in the honeycomb lattice contains two sites depicted
as $A$ and $B$.
We define three unit vectors from site $A$ to its three neighbouring
sites $B$ as
\bea
\hat e_{1,2}=\pm \frac{\sqrt 3}{2} \hat e_x+\frac{1}{2} \hat e_y, \ \ \
\hat e_3=-\hat e_y,
\eea
and their differences $\vec b_i=\frac{1}{2} \epsilon_{ijk} 
(\hat e_j -\hat e_k)$ as
\bea
\vec b_3&=&\sqrt 3 \hat e_x , \ \ \
\vec b_{1,2}=-\frac{\sqrt 3}{2} \hat e_x
\pm \frac{3}{2} \hat e_y.
\eea
The projections of the $p$-orbitals along the $\hat e_{1,2,3}$
directions are defined as
\bea
p_{1,2}= \pm \frac{\sqrt 3}{2} p_x+\frac{1}{2} p_y, \ \ \,
p_{3}=-p_y.
\eea
Only two of them are linearly independent.
In the realistic optical potential depicted in Fig. \ref{fig:potential} A, 
the potential distribution inside each optical site is only 
approximately isotropic in the $xy$-plane.
Away from the center, the potential exhibits a 3-fold rotational
anisotropy.
The  point group symmetry respect to the center of each site
is reduced into $C_{3V}$ including the 3-fold rotation and reflection.
Nevertheless, as required by this symmetry, $p_{x,y}$ remain 
degenerate and each of $p_{1,2,3}$ defined
above is still the on-site eigenstates with the
orientation along the corresponding bond direction.
But they are no longer purely parity odd due to the breaking of
the inversion symmetry respect to the center of each optical site.
(The overall inversion symmetry respect to the
center of the each potential maximum is still preserved, but this
involves the transformation among different sites.)

The $\sigma$-bonding part in the kinetic energy reads
\bea
H_{0}&=&t_\pp\sum_{\vec r \in A, i=1\sim 3} 
\{ p^\dagger_{\vec r,i} p^{\vphantom\dagger}_{\vec r + a \hat e_i,i}
+h.c. \} -\mu \sum_{\vec r \in A
  \oplus B} n_{\vec r}, \ \ \, \ \ \,
\label{eq:ham0}
\eea 
where the summation over $\vec r$ in the first term is only 
on the $A$ sublattice,
$a$ is the nearest neighbor distance, and $n_{\vec r}=n_
{\vec r,x}+n_{\vec r,y}$ is the total particle number 
in both $p_x$ and $p_y$ orbitals
at the site $\vec r$.
$t_\pp$ is positive due to the dominant odd parity component
of the $p$-orbitals  and is set to 1 below.
Eq. \ref{eq:ham0} neglects the much smaller $\pi$-bonding $t_\perp$
terms which in principle exist, and their effects will be
discussed in Sect. \ref{subsect:pibond}.

Next we discuss the spectrum of the tight-binding Hamiltonian 
Eq. \ref{eq:ham0}.
In momentum space, we define a four-component spinor as
\bea
\psi(\vec k)=(p_{Ax}(\vec k),p_{Ay}(\vec k), 
p_{Bx}(\vec k),p_{By}(\vec k))^T,
\eea
where each component is the Fourier transform of the $p_{x,y}$-orbit
in site $A$ or $B$.
Then Eq. \ref{eq:ham0} becomes
\bea
H_0&=&t_\pp\sum_k \psi^\dagger_\alpha(\vec k)
\Big\{ H_{\alpha\beta}(\vec k) -\mu \delta_{\alpha\beta} \Big\}
\psi_\beta(\vec k),
\eea
where the matrix kernel $H_{\alpha\beta}(\vec k)$ takes the structure as
{\small
\bea
\left(
\begin{array}{cccc}
0&0& \frac{3}{4} (e^{i \vec k \cdot \vec e_1} +e^{i \vec k \cdot \vec e_2})
&\frac{\sqrt 3}{4} (e^{i \vec k \cdot \vec e_1} -e^{i \vec k \cdot \vec e_2})
\\
0&0& \frac{\sqrt 3}{4} (e^{i \vec k \cdot \vec e_1} -e^{i \vec k
\cdot \vec e_2})&
\frac{1}{4} (e^{i \vec k \cdot \vec e_1} +e^{i \vec k \cdot \vec e_2})
+e^{i \vec k \cdot \vec e_3}\\
h.c.& & 0&0 \\
   & & 0&0
\end{array}
\right). \nn 
\eea
}

Its spectrum is symmetric respective to zero because the sign of the $t_\pp$ 
term can be flipped by changing the sign of the $p_{x,y}$-orbitals in one 
sublattice but not the other.
The dispersion relations of the four bands read
\bea
E_{1,4}&=&\mp \frac{3}{2} t_\pp,  \ \ \
E_{2,3}= \mp \frac{t_\pp}{2} \sqrt{3+2\sum_i \cos \vec k\cdot \vec b_i}
\eea
as shown in Fig. \ref{fig:hnycmb_band}.
Interestingly, the band structure exhibits two flat bands
$E_{1,4}$ over the entire 2D Brillouin zone.
The corresponding eigenvectors can be found analytically as
\bea
\psi_{1,4} (\vec k)&=&\frac{1}{\sqrt {N_0(\vec k)}}
\Big\{ \frac{1}{\sqrt 3}[f_{23}^*(\vec k)-f_{31}^*(\vec k)], \ \ \
-f_{12}^*(\vec k), \nn \\ \nn\\
&&\pm\frac{1}{\sqrt 3}[f_{23}(\vec k)-f_{31}(\vec k)], \ \ \
\mp f_{12}(\vec k)
\Big\}^T,
\label{eq:eigenvectors}
\eea
where $f_{ij}=e^{i\vec k \cdot \hat e_i}-e^{i\vec k \cdot \hat e_j}$
and the normalization factor reads
\bea
N_0(\vec k)&=&\frac{8}{3}(3-\sum_{i} \cos \vec k \cdot \vec b_i).
\eea
On the other hand, the $E_{2,3}$ bands are dispersive exhibiting
the Dirac cone structure, whose band width is determined by $t_\pp$.
We construct a new set of basis which are orthogonal to  
$\psi_{1,4}(\vec k)$ and span the subspace for the $E_{2,3}$ bands
as
\bea
\phi(\vec k)&=&\sqrt{\frac{2}{N_0}}\Big\{f_{12}(\vec k),\ \ \ 
\frac{1}{\sqrt 3}
(f_{23}(\vec k)-f_{31}(\vec k)),~ 0,~ 0\Big\}, \nn \\
\phi^\prime(\vec k)&=&\sqrt{\frac{2}{N_0}}
\Big\{0,~ 0,~ f^*_{12}(\vec k),\ \ \ \frac{1}{\sqrt 3}
(f^*_{23}(\vec k)-f^*_{31}(\vec k))\Big\}. \nn \\
\eea
Then the Hamiltonian becomes the same as in graphene
\bea
H_{23}(\vec k)=-\frac{t_\pp}{2}\left( \begin{array}{cc}
0& \sum_i e^{-i \vec k \cdot \hat e_i}\\
\sum_i e^{i \vec k \cdot \hat e_i}&0
\end{array}
\right).
\label{eq:dirac}
\eea
Two Dirac cones appear at $K_{1,2}=(\pm\frac{4\pi}{3\sqrt 3 a},0)$.
The eigenvectors of the bands $E_{2,3}(\vec k)$ read
\bea
\psi_{2,3}(\vec k)=\frac{1}{\sqrt 2}
\big\{\phi(\vec k)\pm e^{i\theta_k} \phi^\prime(\vec k) \big\},
\eea
with the angle of $\theta_{\vec k}$ 
\bea
\theta_k=\mbox{arg}(\sum_i e^{i\vec k \cdot \hat e_i}).
\label{eq:angle}
\eea

\subsection{The localized eigenstates}

\begin{figure}
\psfrag{R}{$\vec R$}
\centering\epsfig{file=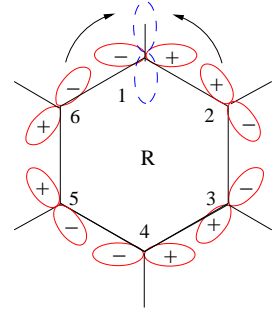,clip=1,width=0.4\linewidth,angle=0}
\caption{
The Wannier-like localized eigenstate for the lowest band. 
The orbital configuration at each site is oriented along 
a direction tangential to the closed loop on which the particle is delocalized.
The absence of the $\pi$-hopping and the destructive interference together
ensure such a state as an eigenstate.
}
\label{fig:wannier}
\end{figure}

The complete flatness of the $E_{1,4}$ bands means that these eigenstates
can be represented as linear superposition of a set of degenerate
\emph{localized} states.
The construction of these localized state is  depicted in  
Fig. \ref{fig:wannier}.
For each hexagon plaquette denoted by its center position $\vec R$,
there exists one such eigenstate for the bottom band $E_1$
\bea
|\psi_{\vec R}\rangle&=&\sum_{j=1}^6 (-)^{j-1}\Big\{ \cos\theta_j
|p_{j,x }\rangle 
-\sin \theta_j |p_{j,y}\rangle \Big\},
\label{eq:wannier}
\eea
where $j$ is the site index and $\theta_j=(j-1)\frac{\pi}{3}$.
The localized eigenstates of the $E_4$ band can be obtained by
flipping the signs of the $p$-orbits on sites $2$, $4$ and $6$
and keep those on sites $1$, $3$ and $5$ unchanged.
The $p$-orbital configuration on each site is perpendicular to the 
links external to the hexagonal loop, thus the $\sigma$-bonding 
forbids the particle to directly hop outside through these links.
Furthermore, the amplitudes for the particle to hop to the $p$-orbital
in the radial direction from the neighbouring sites vanish
due to the destructive interference as shown in Fig. \ref{fig:wannier}.
The particle is trapped in the plaquette without
``leaking'' to outside, and thus $|\psi_{\vec R}\rangle$ is the
eigenstate with the energy of $E_1$.
The states $|\psi_{\vec R}\rangle$ are all
linearly-independent apart from one overall constraint 
$\sum_{\vec R} |\psi_{\vec R}\rangle=0$ under
periodic boundary conditions.
The localized states on two neighbouring edge-sharing plaquettes 
are not orthogonal to each other. 

The Bloch wave states in the flat band $E_1$ are constructed as
\bea
|\psi_{1,k}\rangle=\frac{1}{\sqrt {N_k}} \sum_k e^{i\vec k \cdot \vec R}
|\psi_{\vec R}\rangle \ \ \ (\vec k \neq (0,0)).
\eea
The doubly degenerate eigenstate at $\vec k =(0,0)$
can not be constructed from the above plaquette states.
They are $|\psi_{\vec k=(0,0)}\rangle_{1,2}=
\sum_{\vec r\in A} |p_{x(y), \vec r} \rangle-
\sum_{\vec r\in B} |p_{x(y), \vec r} \rangle$.

\subsection{Orbital configuration at $\vec k=(0,0)$ and $K_{1,2}$}
The major difference between the physics of $p_{x,y}$-orbital bands
and that of graphene is the orbital degree of freedom.
The orbital configuration for each band varies as
lattice momentum $\vec k$ changes in the Brillouin zone.
Around the center of the Brillouin zone $\vec k=(0,0)$, 
the Hamiltonian can be expanded as
\bea
H=\frac{3}{2} \tau_1\otimes I -\frac{3}{4} k_y \tau_2 \otimes\sigma_3
-\frac{3}{4} k_x \tau_2\otimes \sigma_1,
\eea
where Pauli matrices $\sigma_{1,2,3}$ describe the $p_{x,y}$-orbital
degrees of freedom, and $\tau_{1,2,3}$ describes
the sublattices $A, B$ degrees of freedom.
The eigenvectors of $\psi_{1,2,3,4}$ can be approximated as
\bea
\psi_{1,4}(\vec k)&=&\frac{1}{\sqrt 2 |k|}
\Big\{ -k_y, k_x, \pm k_y, \mp k_x \Big\}, \nn \\
\psi_{2,3 }(\vec k)&=&\frac{1}{\sqrt 2 |k|}
\Big\{ k_x, k_y, \mp k_x, \mp k_y \Big\}.
\eea
Thus around $\vec k=(0,0)$, the orbital configuration is polar-like,
i.e., a real combination of $p_{x,y}$.
The orbital orientation in each site is either parallel or perpendicular
to $\vec k$.

\begin{figure}
\centering\epsfig{file=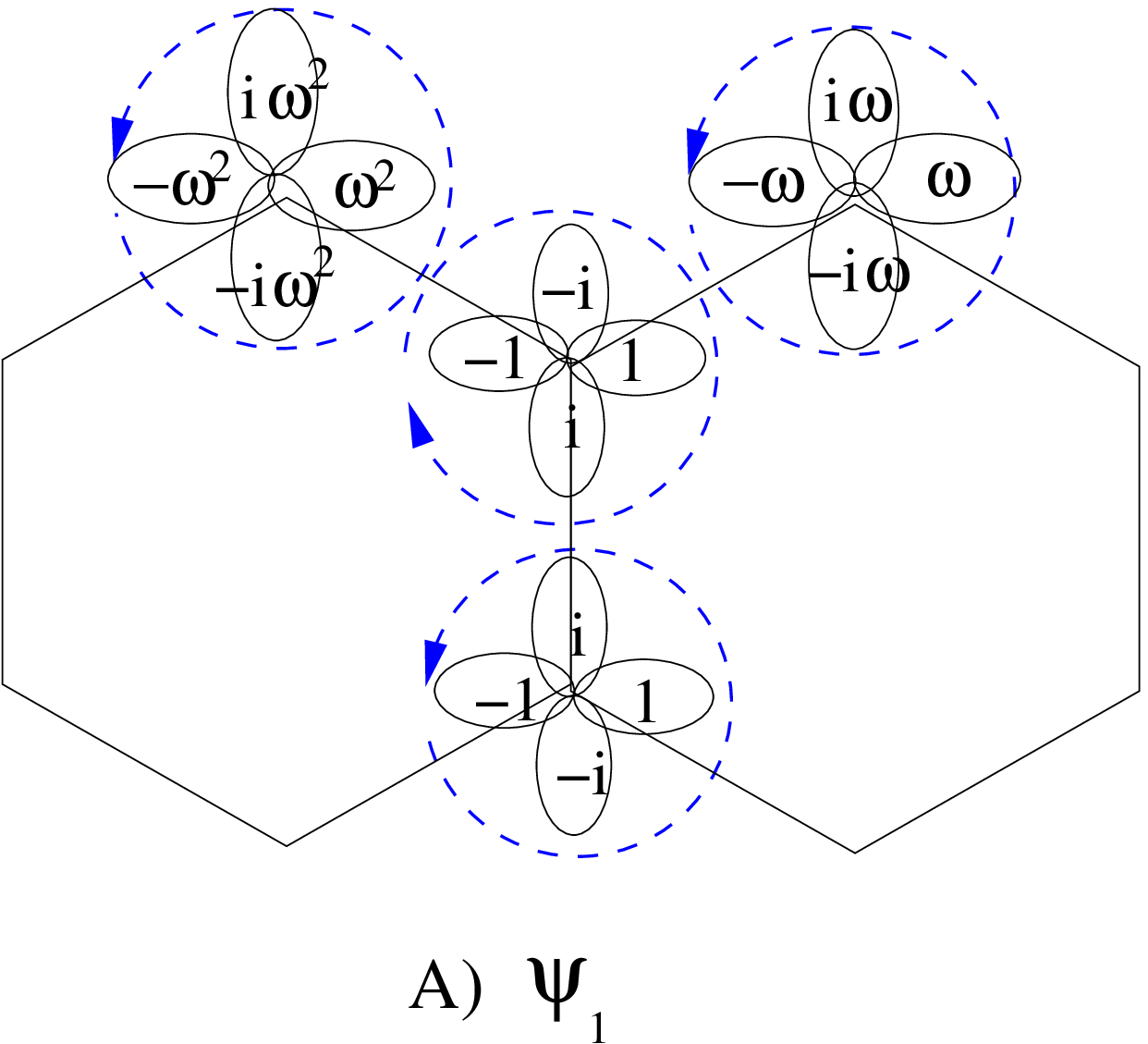,clip=1,width=0.6\linewidth,angle=0}
\centering\epsfig{file=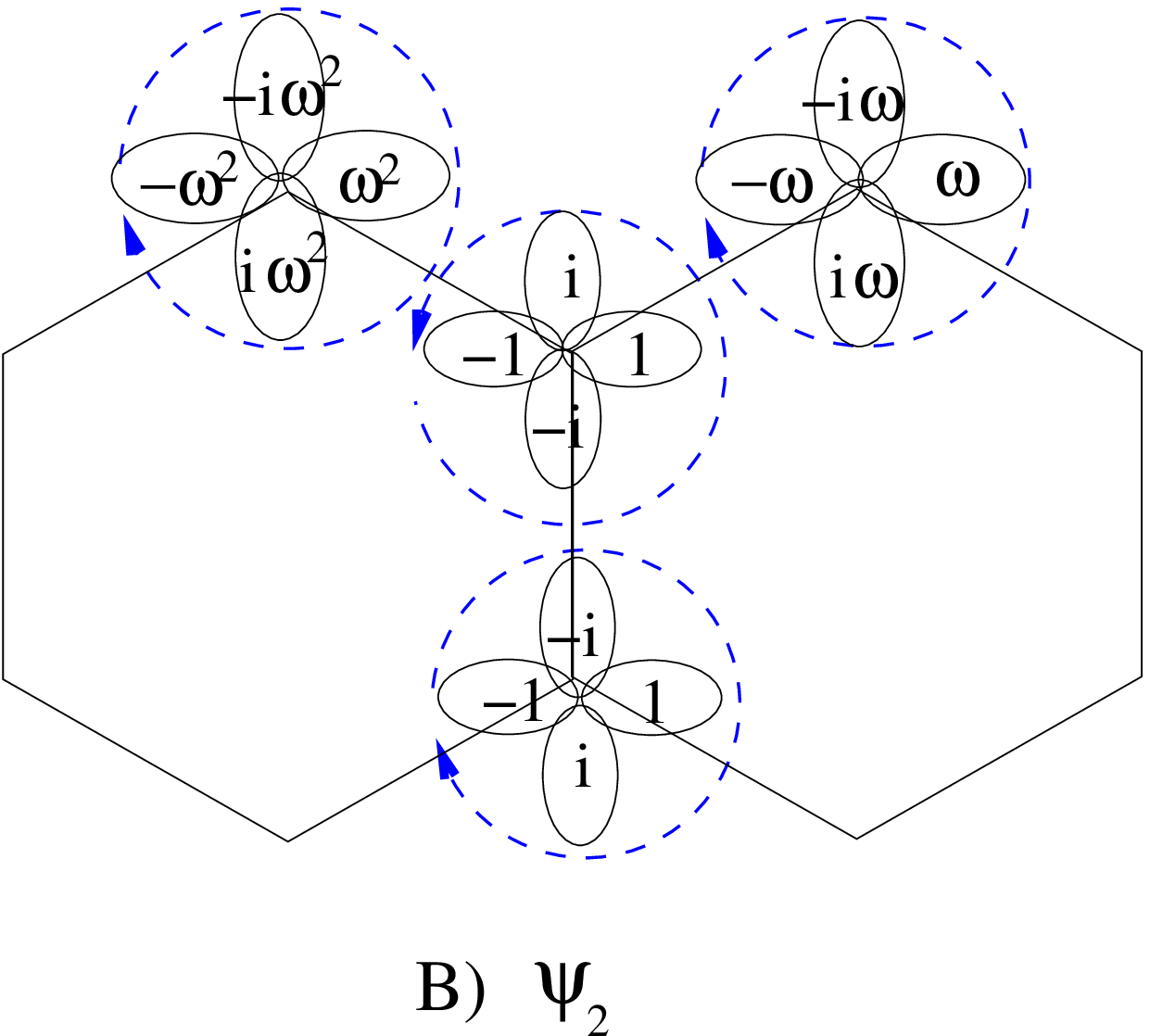,clip=1,width=0.6\linewidth,angle=0}
\caption{The orbital configurations of eigenstates at $K_1$,
which are of $p_x\pm ip_y$ type.
The phase of each lobe  is presented ($\omega=e^{i\frac{2\pi}{3}}$).
A) $\psi_1(\vec K_1)$ ($E_1=-\frac{3}{2} t_\pp$).
B) $\psi_2(\vec K_1)$ with $\alpha_k=0$ ($E_2=0$). 
}
\label{fig:hnycmb_obcf}
\end{figure}

Now let us investigate the orbital configurations 
around the vertices of  $K_{1,2}$ of the Brillouin zone.
Around $K_1$, the Hamiltonian can be expanded as
\bea
H(\vec k) &=& -\frac{3}{4} \Delta k_x \tau_1 \otimes I
+\frac{3}{4} \Delta k_y ~ \tau_2 \otimes I \nn \\
&-&(\frac{3}{4}+\frac{3}{8}\Delta k_x)~ \tau_1 \otimes \sigma_3 
-\frac{3}{8} \Delta k_y~ \tau_2 \otimes \sigma_3  \nn \\
&-&\frac{3}{8} \Delta k_y ~\tau_1 \otimes \sigma_1 
-(\frac{3}{4}-\frac{3}{8} \Delta k_x) \tau_2 \otimes \sigma_1,
\eea
where $\Delta \vec k=\vec k-\vec K_1$ and
$g_{\pm}(\vec k)=\Delta k_x \pm i \Delta k_y.$
The eigenvectors of the flat bands $E_{1,4}$ can be approximated as
\bea
\psi_{1,4} (\vec k)&=&\frac{1}{2}\Big\{ 1+\frac{g_{+} (\vec k)}{2}, ~
-i [1-\frac{g_{+}(\vec k)} {2} ], \nn \\
&\pm& [1+\frac{g_{-}(\vec k)}{2}], 
~\pm  i [1-\frac{g_{-}(\vec k)}{2}]  \Big\}^T.
\eea
Similarly, the eigenvectors of the dispersive bands
$E_{2,3}$ are approximated as
\bea
\psi_{2,3}(\vec k)&=&\frac{1}{ 2} \big\{1-\frac{g_{-}(\vec k)}{2},~
i (1+\frac{g_-(\vec k)}{2}), \nn \\
&&\pm e^{i\alpha_k} (1-\frac{g_+(\vec k)}{2},~
-i (1+\frac{g_+(\vec k)}{2}) \big\}^T, \ \ \
\eea
where $\alpha_k$ is the angle defined in Eq. \ref{eq:angle}.
Thus the orbital configuration at $\vec k =\vec K_1$ 
on each site is the axial state $p_x\pm i p_y$ as depicted
in Fig. \ref{fig:hnycmb_obcf}.
This is in contrast to the polar configuration at $\vec k=(0,0)$.
The orbital configuration at $\vec k = \vec K_2$ can be obtained
by performing time reversal transformation.

\subsection{$\pi$-bonding term and other perturbations}
\label{subsect:pibond}

\begin{figure}
\centering\epsfig{file=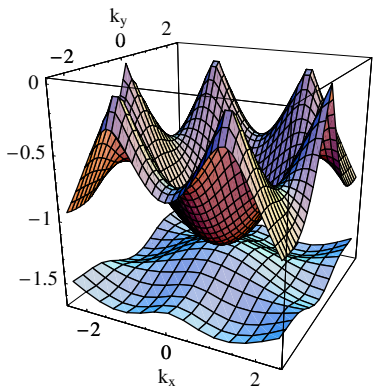,clip=1,width=0.6\linewidth,angle=0}
\caption{Dispersion of the two-lowest $p_{x,y}$-orbital bands
  $E_{1,2}$ in the presence of the $\pi$-bonding $t_\perp$.
  The bottom band is no longer rigorously flat but acquires a narrow width
  of $t_\perp$. 
}
\label{fig:hnycmb_band2}
\end{figure}

\begin{figure}
\centering\epsfig{file=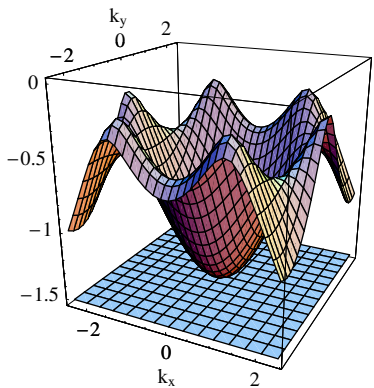,clip=1,width=0.6\linewidth,angle=0}
\caption{Dispersion of the two-lowest $p_{x,y}$-orbital bands
  $E_{1,2}$ in the presence of the on-site potential.
}
\label{fig:hnycmb_band3}
\end{figure}

The $\pi$-bonding term in principle exists in the realistic optical
lattices.
We define the projections of $p_{x,y}$-orbitals perpendicular to the
$\hat e_{1,2,3}$ directions as
\bea
p^\prime_{1,2}=-\frac{1}{2} p_x \pm \frac{\sqrt 3}{2} p_y, \ \ \
p^\prime_3=p_x.
\eea
The $\pi$-bonding term can be written as
\bea
H_\pi&=&-t_\perp\sum_{\vec r \in A, i=1\sim 3} 
\Big \{ p^{\prime \dagger}_{\vec r,i} p^\prime_{\vec r + a \hat e_i,i}
+h.c. \Big \}.
\label{eq:hampi}
\eea 
Please note that the hopping integral of the $\pi$-bonding has
the opposite sign to that of the $\sigma$-bonding.
In this case, the bottom and top bands $E_{1,4}$ are no longer rigorously
flat but develops a narrow width $3 t_\perp$ as depicted in
Fig. \ref{fig:hnycmb_band2}.
The $E_1$ and $E_2$ bands still touch at the center of the Brillouin
zone.
This can be understood from the structure of the localized eigenstates
in Fig. \ref{fig:wannier}. 
The $\pi$-bonding term causes the particle 
leaking off the plaquette and thus correspondingly develops the band width.
Nevertheless we will show in Sect. \ref{subsect:continuum} 
that in the realistic optical potential
that such an effect is negligibly small.


Next we discuss the case that the $A$ and $B$ sites are with different 
on-site potentials.
In the graphene-like systems, this corresponds to a mass term in 
the Dirac point.
In the $p_{x,y}$-orbital systems, such a term can be described as
\bea
H&=& \Delta E \Big\{ \sum_{\vec r \in A} n_{\vec r}-
\sum_{\vec r \in B} n_{\vec r} \Big\}.
\label{eq:hammass}
\eea 
The spectrum is depicted in Fig. \ref{fig:hnycmb_band3} with the
opening of a gap of $\Delta E$ in the Dirac points as usual.
Interestingly, the flat band feature remains unchanged.
This can be understood in terms of the localized eigenstate
picture.
In this case, the localized eigenstates of the $E_1$ band 
still possess a similar configuration as in Fig. \ref{fig:wannier},
but their wavefunctions distribute with different 
 weights on $A$ and $B$ sublattices.

\subsection{Band structure from the continuum optical potential}
\label{subsect:continuum}

\begin{figure}
\centering\epsfig{file=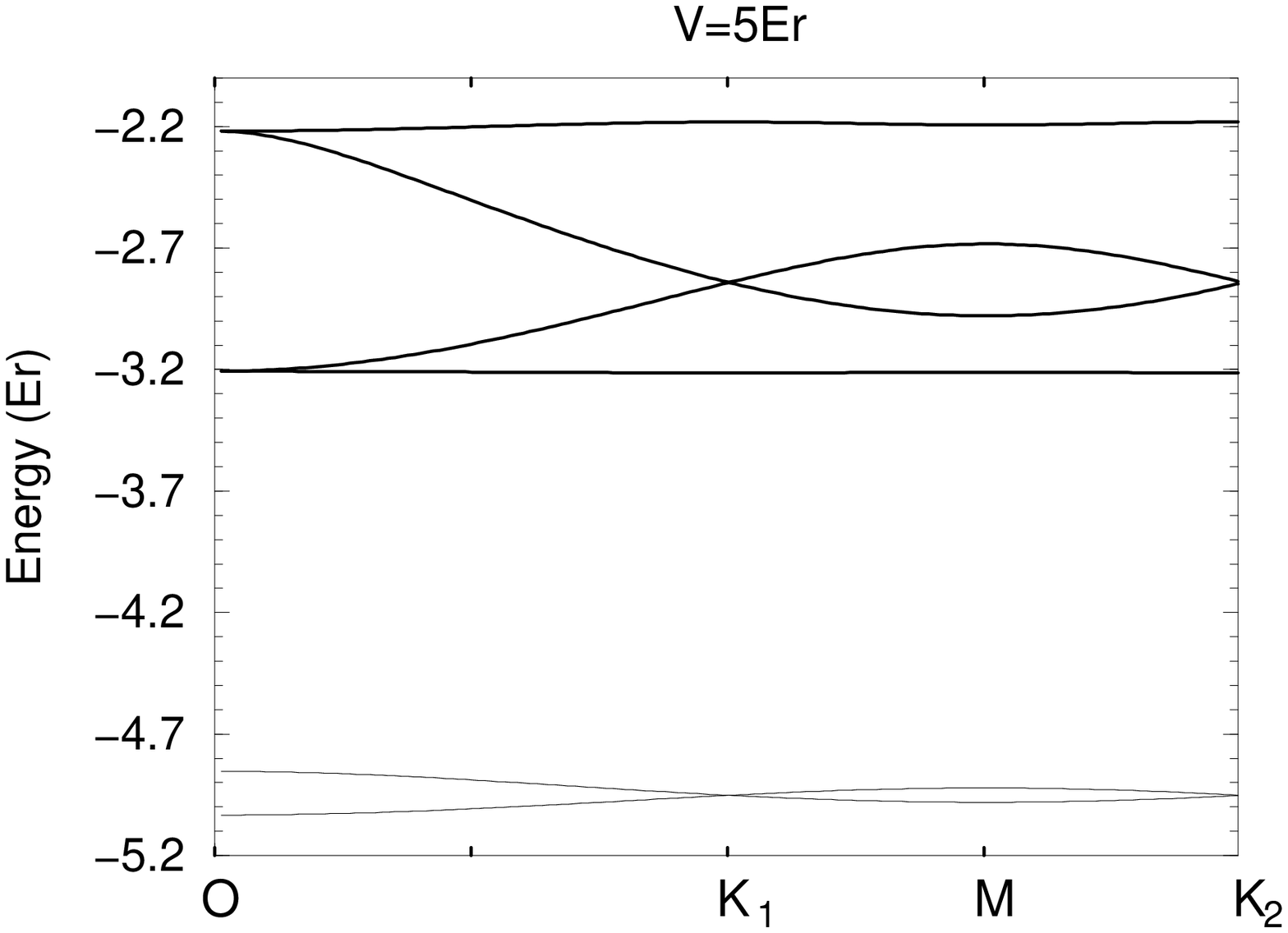,clip=1,
width=0.8\linewidth,angle=0}

\ \ \

\centering\epsfig{file=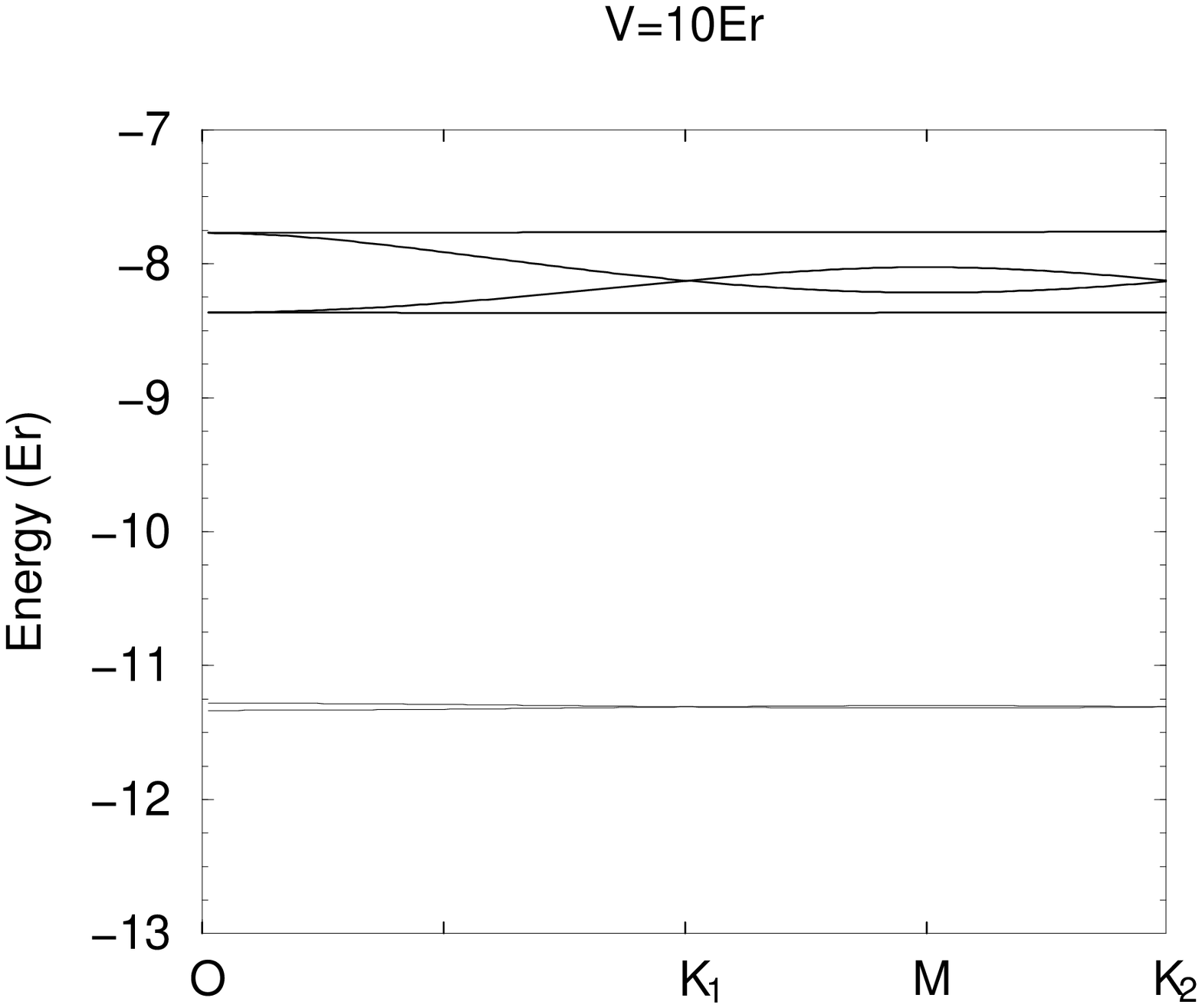,clip=1,
width=0.8\linewidth,angle=0}
\caption{Band structures for the realistic optical potential in
Eq. \ref{eq:potential} along the path from $O \rightarrow 
K_1 \rightarrow M \rightarrow K_2$ in the Brillouin zone with
A) $V/E_r=5$ and B) $V/E_r=10$.
\label{fig:hnycmb_englvl}
}
\end{figure}

\begin{figure}
\centering\epsfig{file=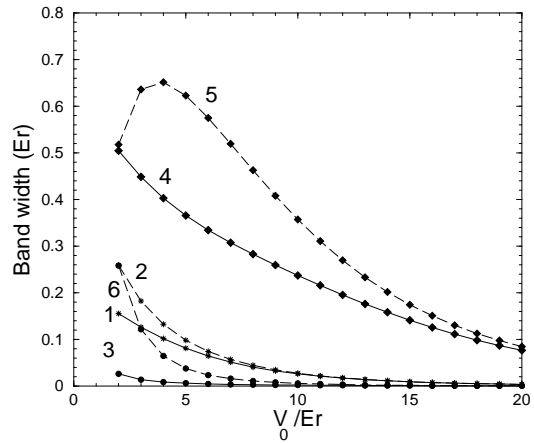,clip=1,
width=0.8\linewidth,angle=0}
\caption{Width of the beginning six bands of the optical potential Eq.
\ref{eq:potential}. Bands 1 and 2 are the $s$-orbital bands. And
bands 3,4,5,and 6 are the $p_{x,y}$ orbital bands, where
those of $3$ and $6$ are flat as discussed before.
}
\label{fig:bandwidth}
\end{figure} 

We numerically calculate the band structure in the realistic 
optical potential of Eq. \ref{eq:potential}.
The band Hamiltonian  becomes
\bea
H=-\frac{\hbar ^2 \nabla^2_r}{2m}+V_0 \sum_{i=1\sim3} 
\cos (\vec p_i \cdot \vec r).
\eea
Since this  is a non-singular sinusoidal potential,
we use the plane wave basis to calculate the matrix elements 
$\avg{\vec k|H|\vec k^\prime}$ where $\vec k^\prime=\vec k\pm 
\vec p_i (i=1\sim 3)$.
For each $\vec k$ in the Brillouin zone, we truncate the matrix
up to $120$ plane-wave basis, which should be sufficient
for  the lowest several  bands.

The band dispersions along the path from $O$ to $K_1$, $M$ and $K_2$
are depicted in Fig. \ref{fig:hnycmb_englvl}.
The locations of $O$, $K_{1,2}$, $M$ in the Brillouin zone
are depicted in Fig. \ref{fig:hnycmb_lat}.
The lowest two bands are of the 
$s$-orbital exhibiting Dirac cones at $K_{1}$ and $K_{2}$.
The next four are of the $p_{x,y}$-orbitals.
The band flatness is largely preserved even with the
realistic optical potential of Eq. \ref{eq:potential}.
In Fig. \ref{fig:hnycmb_englvl} A with $V/E_r=5$,
the bottom one of the four $p$-orbital bands 
is nearly flat with the width of
$7 \times 10^{-3} E_r$ which is only $2\%$ of that of the
second one which is $0.35 E_r$.
The width of the top band is $4\times 10^{-2} E_r$ which is still
small but considerably larger than that of the bottom one.
The third band is the widest one with the width of $0.62 E_r$.
As we can see, the particle-hole symmetry in the tight-binding model 
is no longer kept because of the unavoidable hybridization 
with other bands and long range hoppings.
The spectra become more symmetric with a strong optical potential
as shown in Fig. \ref{fig:hnycmb_englvl} B ($V/E_r=10$)
in which the tight-binding model is a better approximation
and long range hoppings can be neglected.
The widths of the beginning six bands 
(the $s$ and $p_{x,y}$-orbital bands) 
as a function of the $V_0/E_r$ are
depicted in Fig. \ref{fig:bandwidth}.

\section{Interactions in the $p_{x,y}$-orbital spinless fermions }
\label{sect:interaction}
In the following, we will mainly consider interacting spinless fermions
in the $p_{x,y}$-orbital bands in the honeycomb lattices, and
leave the research for spinful fermions in future publications.
The preparation of spinless fermions can be controlled by cooling
the system in the external Zeeman field.
Due to the lack of spin relaxation mechanism in cold atom systems,
the system will remain in the spin polarized state.
The spinless fermions have been realized in many experiments.
In particular, the strongly correlated polarized spinless fermion
systems have been realized by using the $p$-wave Feshbach 
resonance \cite{regal2003,gaebler2007,ticknor2004,zhang2004}.
Therefore, in contrast to solid
state electronic systems, where spin is almost always an important quantum
dynamical variable, the cold atom fermionic systems created by Feshbach
resonance, can be prepared as spinless (i.e. spin polarized), and
consequently, our current spinless theory applies to such systems without
any modifications.  Of course, the problem of creating a laboratory $p_{x,y}$
orbital graphene system in cold atomic gases still remains, but given the
rapid current experimental developments in fermionic cold atom matter, we
are optimistic that our prposed system should soon be realized in practice.

Because of the orbital degeneracy, the on-site interaction for 
spinless fermions remains Hubbard-like 
\bea
H_{int}= U \sum_{\vec r } n_{\vec r, x} n_{\vec r, y},
\label{eq:hamint}
\eea 
where the on-site interaction $U$ is
\bea
U&=&\int d \vec r_1 d\vec r_2 V(\vec r_1-\vec r_2)
\Big( (\psi_{p_x}(\vec r_1) \psi_{p_y}(\vec r_2))^2 \nn \\
&-&(\psi_{p_x}(\vec r_1) \psi_{p_y}(\vec r_1)
\psi_{p_x}(\vec r_2) \psi_{p_y}(\vec r_2)
\Big).
\eea
Due to Paul's exclusion principle, the $s$-wave scattering vanishes, 
and thus the $p$-wave scattering is the leading order contribution
which is typically weak for low energy particles.
The $p$-band fermions have high kinetic energy, and thus
their $p$-wave scattering might not be small.
To enhance $U$, we can use the $p$-wave Feshbach resonances 
among spinless fermions (e.g. $^{40}$K \cite{gaebler2007,ticknor2004}, 
$^6$Li\cite{zhang2004}).
Although we do not want the system staying too close to the
resonance because of the large atom loss rate there, an enhancement
of $U$ to the order of the recoil energy $E_R$ while maintaining
the stability of the system is still reasonable.

Another possible method is to use atoms with large magnetic
moments which interact through magnetic dipole-dipole interactions as
\bea
V(\vec r_1-\vec r_2)=\frac{1}{r^3}
\{ \vec m_1\cdot \vec m_2 - 3 (\vec m_1 \cdot \hat r)
(\vec m_2 \cdot \hat r) \},
\eea
where $r=|\vec r_1-\vec r_2|$ and $\hat r=(\vec r_1-\vec r_2)/r$.
The fermionic atom of $^{53}$Cr is a good candidate whose magnetic moment is
$m_{Cr}=6\mu_B$ (Bohr magneton).
The spin polarization can be controlled by an external magnetic
field. Below we give an estimation of $U$  from the
magnetic dipole interaction.
The vibration frequency in each site can be obtained
as $\omega_{x,y}=\sqrt{\frac{3}{2} V_0 E_r}$. The length
scale of the $p_{x,y}$-orbitals ($l_{x,y}=\sqrt{\hbar/m\omega_{x,y}}$)
is typically one order smaller than $a$.
For example, we estimate that $l_{x,y}/a\approx 0.2$ at
$V_0/E_r=5$.
Assuming strong confinement in the $z$-axis $l_z\ll l_{x,y}$,
the vector $\vec r_1-\vec r_2$ 
linking two atoms in $p_x$ and $p_y$ orbits almost 
lies in the $xy$-plane.
When the fermion spin is polarized along the $z$-axis, 
the interaction is repulsive and $U$ can be approximately estimated as
\bea
U\approx \frac{m_{Cr}^2}{\avg{r^3}}  (1-3\avg{\cos^2 \theta}),
\eea
where $\theta$ is the angle between $\vec r_1-\vec r_2$
and the $z$-axis.
We estimate $\avg{r}=\sqrt{2l_{xy}^2+l_z^2}$ and $\cos\theta=l_z/r$,
and find that $U$ can reach the order of $E_r$.
For example, if we use the laser wavelength $\lambda\approx 0.8 \mu m$,
$V/E_r=5$ (so that $l_{x,y}\approx 0.2 a$) and $l_z=0.2 l_{x,y}$,
we arrive at $U=2.2$ KHz or approximately $100nK$.
Increasing $V/E_r$ can further increase $U$ and suppress $t_\pp$,
thus drive the system into even stronger correlation regime.
$U$ can be adjusted from repulsive to attractive 
by tuning the polarization direction from perpendicular to parallel
to the $xy$-plane.

\section{The Wigner Crystal state at $\frac{1}{6}$-filling}
\label{sect:1/6}
In this section, we discuss interaction effects in the
$p_{x,y}$-orbital systems.
When the flat band is partially filled, interaction effects
dominate the physics.
In particular, a Wigner crystal state is stabilized even with
the shortest range on-site interaction.
We will mainly study the spinless fermion system below, and
also  give a brief discussion on the boson systems, 
but leave the study of the spinful fermion
systems to a future publication.

\subsection{Close packed plaquette state}
\begin{figure}
\psfrag{R} {$\vec R$}
\psfrag{R1}{$\vec R_1$}
\psfrag{R2}{$\vec R_2$}
\psfrag{R3}{$\vec R_3$}
\centering\epsfig{file=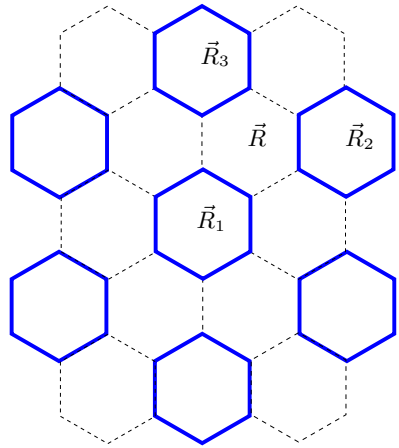,clip=1,width=0.6\linewidth,angle=0}
\caption{ The configuration of the close-packed Wigner-crystal
state for both bosons and fermions at $n=1/6$. 
Each thickened plaquette has the same
configuration as in Fig. \ref{fig:wannier}. 
From Wu {\it et al} \cite{wu2007}.
}
\label{fig:closepack}
\end{figure}

Due to the complete suppression of the kinetic energy in the flat band,
the effect of interactions is non-perturbative 
when the flat band is partially filled.
Interestingly, at sufficiently low particle density $n\leq \frac{1}{6}$, 
the exact many-body ground state can be easily constructed as follows.
Each individual particle localizes into a plaquette state depicted in 
Fig. \ref{fig:wannier}.
Any arrangement of these plaquette states avoiding touching each other
is the kinetic energy ground state and costs zero interaction energy.
Since the interaction is repulsive, this class of states also 
minimize the interaction energy and thus
they constitute the many-body ground states.
If we fix the particle density at $n<1/6$, the ground state configurations 
have large degeneracy corresponding to all the possible ways to arrange 
these hard hexagons.

Another class of systems exhibiting similar behavior is the frustrated 
magnets near full polarization in a large external magnetic field.
The Holstein-Primakoff magnons, which are bosons, have
a dispersionless flat band over the magnetic Brillouin zone.
Interactions among magnons  result in the magnon crystal state 
and magnetization plateau \cite{zhitomirsky2004} near the full 
polarization. 
However, this flat band behavior is difficult to  observe because 
a very strong magnetic field to drive the system
close to the full polarization is required.
This means that the Zeeman energy reaches the exchange energy $J$
which is typically larger the order of meV.
Flat band phenomenon also appears in systems of  ``fermion condensation''
where strong interactions drive an originally dispersive band to 
flat within a finite width around the Fermi energy  \cite{khodel1990}.
This has been proposed to explain the Curie's law behavior of the
magnetic susceptibility in the itinerant  heavy fermion compound 
CeCoIn$_5$ system \cite{khodel2005}.

The close packed plaquette pattern without overlapping each other
is depicted in Fig. \ref{fig:closepack} corresponding to the
filling of $n=\frac{1}{6}$.
The completely filled lowest flat band corresponds to $n=\frac{1}{2}$,
thus this close packed plaquette pattern corresponds to
$\frac{1}{3}$-filling of the flat band.
This state breaks the lattice translational symmetry and is
three-fold degenerate. 
The other two equivalent states can be obtained by translating
the state in Fig. \ref{fig:closepack} along $x$-axis in the 
right or left direction at one lattice constant.

\subsection{Stability of the Wigner crystal state}
\begin{figure}
\centering\epsfig{file=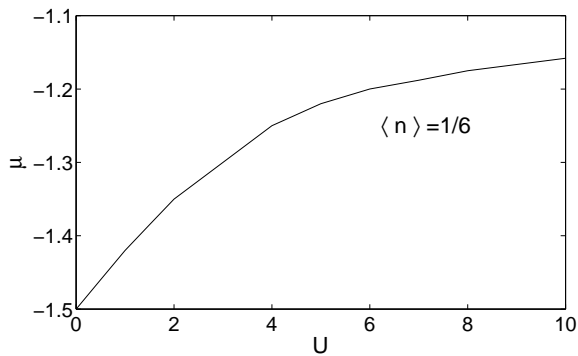,clip=1,width=0.9\linewidth,angle=0}
\caption{The phase boundary of the incompressible plaquette Wigner 
crystal state of spinless fermions at $\avg{n}=\frac{1}{6}$. 
From Wu {\it et al} \cite{wu2007}.
}\label{fig:gap}
\end{figure}

The above Wigner crystal state is a gapped state. We can give
a rough estimation for
a upper limit of the charge gap by constructing a trial wavefunciton
for putting an extra particle in the close packed state in 
Fig. \ref{fig:closepack}.
In the weak interaction case $(U/t_\pp\ll 1)$, we can put the
extra atom in the plaquette state located at $\vec R$ which is 
adjacent to three occupied plaquettes $\vec R_{1,2,3}$.
Since there is already $1/6$ atom on average per site, the cost
of the repulsion is $\frac{U}{6}$.
On the other hand, in the strong coupling case  $(U/t_\pp \gg 1)$,
we put the particle into an excited state of the occupied 
plaquette $\vec R_1$ while fixing the orbital configuration on each site. 
Because fermions are spinless, the cost of energy comes from
the kinetic part with the value of $\frac{3}{4}t_\pp$.
Thus we obtain the upper limit for the charge gap 
which is determined by interaction at small values of $U$ and by kinetic 
energy $t_\pp$ at large values of $U$ as
\bea
\Delta <  \mbox{min}(\frac{1}{6} U, \frac{3}{4}t_\pp).
\eea
The above intuitive picture can be make more rigorous by performing
self-consistent mean field treatment to the Hamiltonian described below.

We decouple Eq. \ref{eq:hamint} both in the direct and exchange channels as
\bea
H_{mf,int}&=& U \sum_{\vec r \in A \oplus B}
\Big\{ n_{\vec r, x} \Big[\frac{\avg{n_{\vec r}}}{2} -\avg {n_{\vec r,1}} \Big]
+n_{\vec r, y}\Big[\frac{\avg{n_{\vec r}}}{2} \nn \\
&&\hspace{-10mm} +\avg {n_{\vec r,1}}\Big]
-p^\dagger_{\vec r x} p^{\vphantom\dagger}_{\vec r y} \Big[\avg {n_{\vec r,2}}
-i\avg {n_{\vec r,3}}\Big]
-h.c. \Big\}, 
\label{eq:mft}
\\
n_{\vec r, 1} &=& \frac{1}{2} (p^\dagger_{\vec r x}p_{\vec r x}-
p^\dagger_{\vec r y}p_{\vec r y}), \nn \\
n_{\vec r, 2} &=&\frac{1}{2} (p^\dagger_{\vec r x} p_{\vec r y}+h.c.), \nn \\
n_{\vec r, 3}&=&\frac{1}{2i} (p^\dagger_{\vec r x} p_{\vec r y}-h.c.),
\eea
where $n_{1,2,3}$ are the pseudo-spin operators.
$n_{\vec{r},1},n_{\vec{r},2}$  are time-reversal invariant,
and describe the preferential occupation of a ``dumbbell-shaped'' 
real $p$-orbital orientation; $n_{\vec{r},3}$ is the
orbital angular momentum, and is time-reversal odd.
We perform a self-consistent mean field solution to Eq. \ref{eq:mft}
plus Eq. \ref{eq:ham0} for the filling level in the range of
$\avg{n}=0\sim 1$.
The self-consistent equation reads
\bea
\avg{n_{\vec{r},i}}=\avg{\Omega[ \avg{n_{\vec{r},j}}] ~| n_{\vec{r},i}|~
\Omega[\avg{n_{\vec{r},j}}] } ~(i,j=1\sim 3),
\eea
where $| \Omega[n_{\vec{r},j}] \rangle$ is the mean field ground state
with the specified configuration of $\avg{n_{\vec r,j}}$.

At the mean field level, we found that $\langle n_{\vec{r},3}\rangle$ 
is zero which means the time reversal symmetry is kept.
We need to take an enlarged unit cell to allow the spatial 
variation of the order parameters.
In order to obtain the plaquette order in Fig. \ref{fig:closepack}, 
this enlarged unit cell covers six sites around a plaquette.
We present the range of chemical potential $\mu$ for the $\frac{1}{6}$
-Wigner crystal state in Fig. \ref{fig:gap}, which corresponds to the
excitation gap.
The charge gap grows roughly linearly with $U$ in the weak interaction
regime, and saturates at a value comparable to $t_\pp$ in the strong 
interaction regime.
Both agree with the above variational analysis.

Next we discuss the effect of the $\pi$-bonding term to the 
$\frac{1}{6}$-state.
Such a term gives a width determined by $t_\perp$ to the originally flat band.
Because the $\frac{1}{6}$-state is gapped, it should remain stable
if the band is sufficiently narrow.
The plaquette state costs the kinetic energy at the
order of $t_\perp$ per particle while it saves the repulsive
interaction at the order of $\frac{U}{6}$.
Thus for small values of $t_\perp$, a stability condition of this state 
can therefore be roughly estimated as $U>6 t_\perp$.
We have checked this  numerically.
For example, setting the $t_\perp/t_\pp=0.1$, we find that
the $\frac{1}{6}$-state survives $U>t_\pp$.
In realistic systems, the ratio of $t_\perp/t_\pp$ is much smaller
than 0.1 with reasonable values of $V/E_r$ as shown in 
Fig. \ref{fig:bandwidth}, thus the 1/6-state can be stabilized
at much smaller values of $U$.

\subsection{Bosonic Wigner crystal state}
In the above hard hexagon state at $n=\frac{1}{6}$, particles are 
separated from each other, thus  particle statistics do not play any role. 
Such a Wigner crystal state should also occur with bosons or 
Bose-Fermi mixtures with repulsive interactions.
The on-site interaction for $p$-band bosons reads
\bea
H_{int}=\frac{U}{2} \sum_{\vec r}
\Big\{ n^2_{\vec r}-\frac{1}{3} L_{z,\vec r}^2\Big\},
\eea   
where $n$ is the total particle number and $L_z$ is the orbital angular
momentum \cite{liu2006,wu2006}. 
The $p$-band bosonic systems have been created experimentally 
\cite{mueller2007}.
The life time of the $p$-band bosons is significantly enhanced
when the particle density per site is less than one, which can be
hundreds of times longer than the hopping timer from one site
to its neighbours.
Thus the $1/6$-state is also experimentally feasible in bosonic systems.

\section{Charge and bond orderings at commensurate
fillings of $n>\frac{1}{6}$}
\label{sect:comm}
In this section, we investigate the charge and bond orderings
at commensurate fillings higher than $\frac{1}{6}$
by using the mean field theory to solve the interacting 
Hamiltonian self-consistently.
We will present the result in both weak and strong coupling regimes,
but leave the detailed investigate of the physics of 
orbital exchange at $n=1$ to a future research. 
In the following calculation, we confine ourselves to the
unit cell up to 6-sites.

\subsection{Weak coupling regime}
\begin{figure}
\centering\epsfig{file=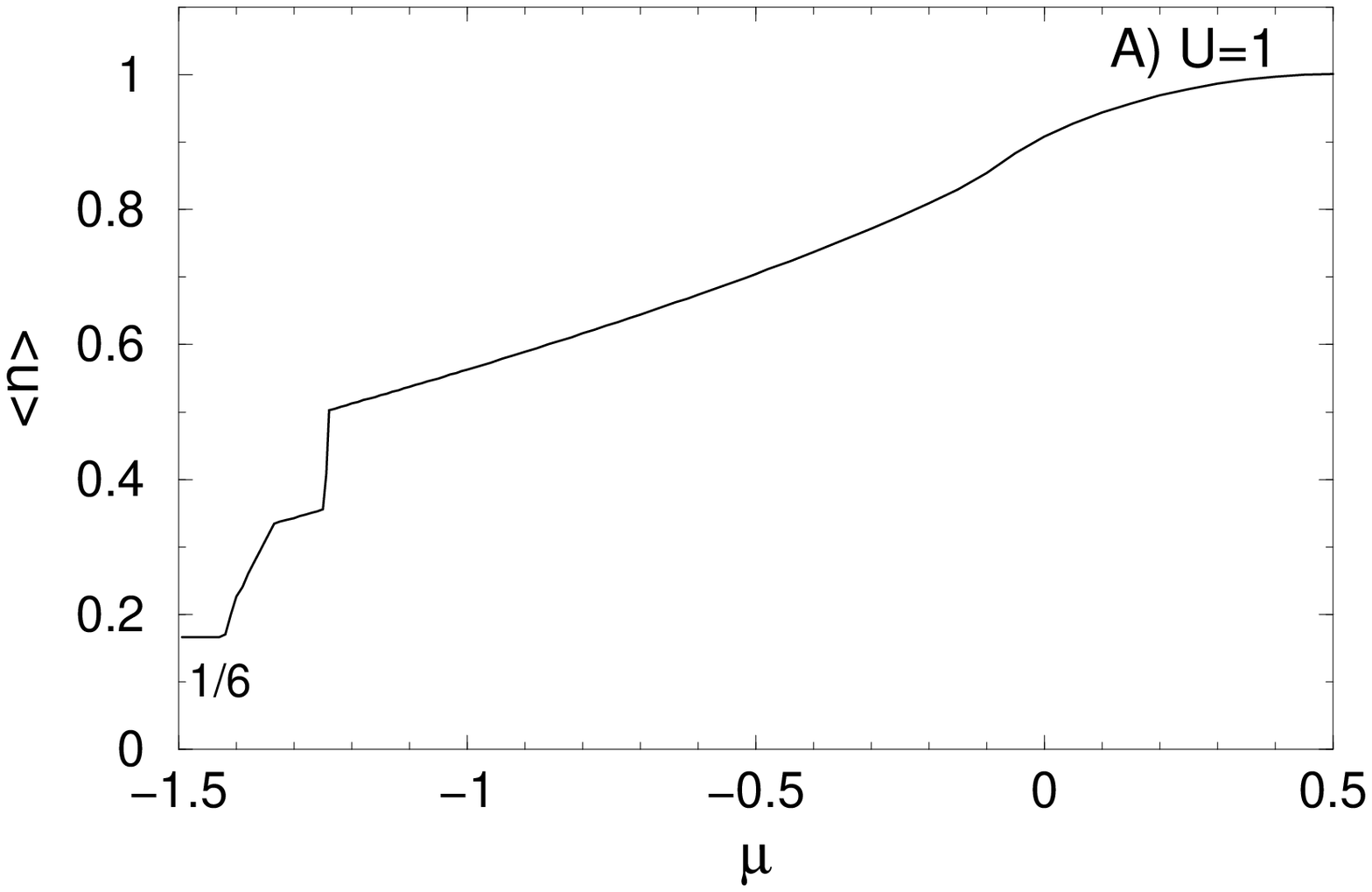,clip=1,width=\linewidth,angle=0}
\centering\epsfig{file=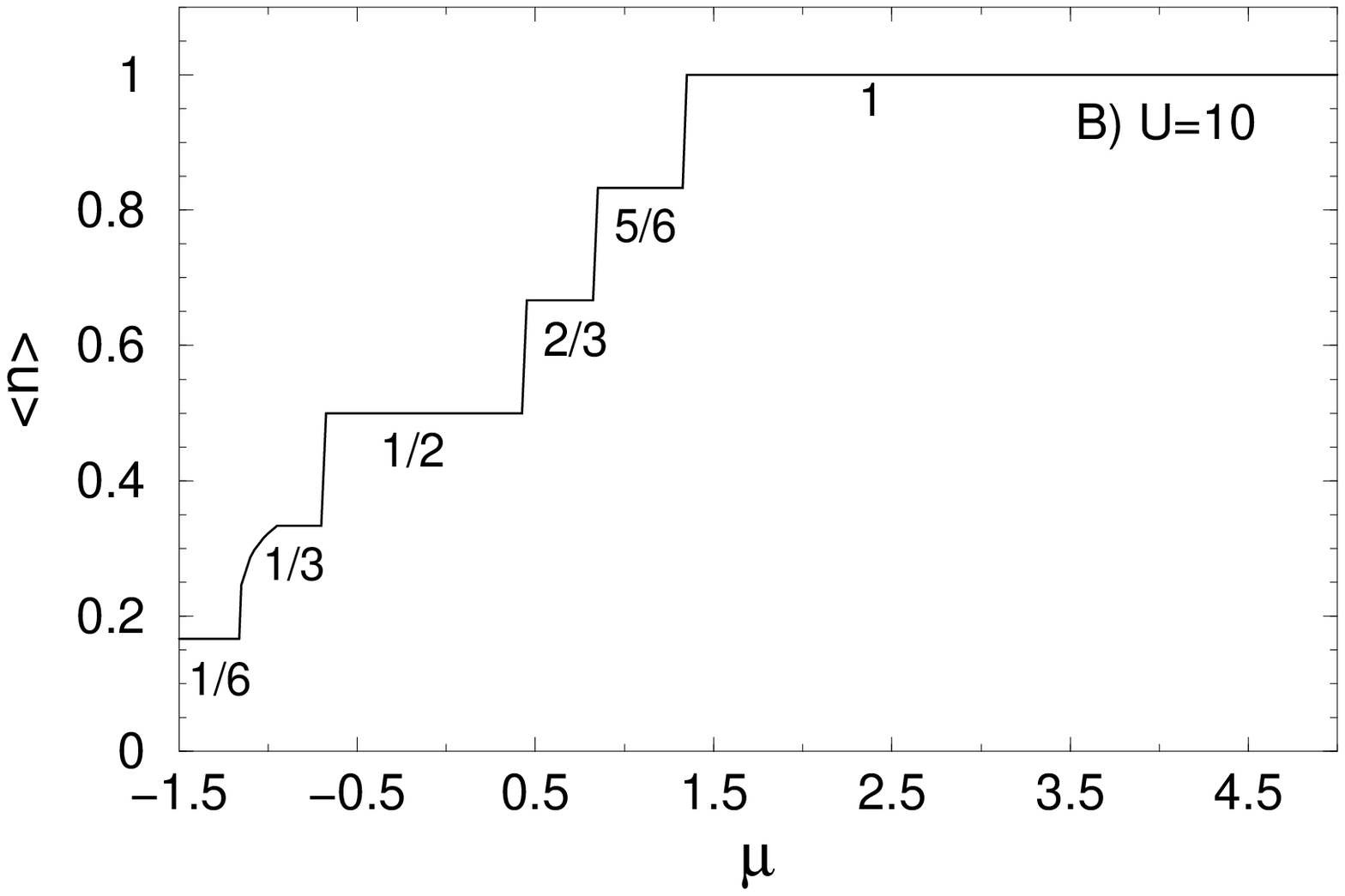,clip=1,width=\linewidth,angle=0}
\caption{The filling $\avg{n}$ vs. the chemical potential $\mu$
for spinless fermions for weak A) and strong B) interactions.
Due to the particle-hole symmetry, only the part with
$\mu$ from the band bottom $-\frac{3}{2}t_\pp$ to $U/2$ is shown.
Only one plateau appears in A)  at $n=\frac{1}{6}$,
while a series of plateaus appear in B)
at $n=1/6, 1/3, 1/2, 2/3, 5/6, 1$. From Wu {\it et al} \cite{wu2007}.
}\label{fig:filling}
\end{figure}  

\begin{figure}
\centering\epsfig{file=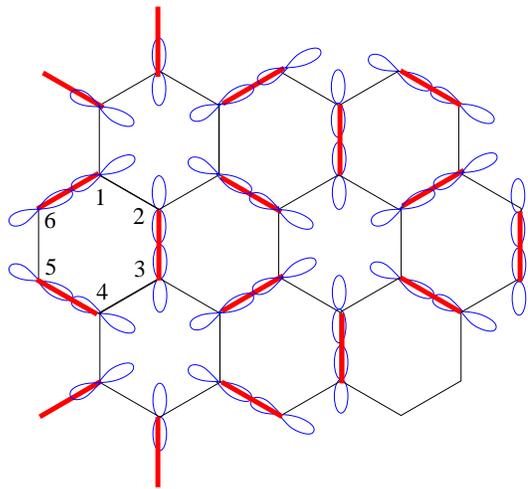,clip=1,width=0.8\linewidth,angle=0}
\caption{Bonding strength dimerization can occur at both $\avg{n}=
1/3$ in the weak coupling regime and $\avg{n}=1/2$ in the strong coupling
regime as depicted by the thickened (red) bonds.
The orbital orentation in the dimer is along the bond direction.
In the weak coupling case ($\avg{n}=1/3$), the the thickened bonds 
correspond to the shared edges of two neighboring plaquette states
in the flat band.
In the strong coupling case ($\avg{n}=1/2$), each dimer contains one particle
as an entangled state of occupied and empty sites.
 }\label{fig:dimer}
\end{figure}

When the filling $n>1/6$, exact solutions are no longer available.
Again we perform the self-consistent mean field solution to the interacting
Hamiltonian.
In the weak coupling regime ($U/t_\pp=1$), we plot the relation of
the filling $\avg{n}$ vs. $\mu$ in Fig.  \ref{fig:filling} A.
As $\mu$ passes the charge gap, the system enters a compressible state.
$\avg{n}$ increases with $\mu$ quickly with a finite but large slope.
This means that particles fill in other states in the flat band.
Due the preexisting crystalline ordered  background,
these states are no longer exactly flat and develop weak dispersions.
This corresponds to adding additional $\frac{N}{6}$ particles
into the background  of the $\frac{1}{6}$-state as depicted 
in Fig. \ref{fig:closepack} ($N$: the total number of lattice sites).
Roughly speaking, these new particles also go into the localized plaquette
states.
When $\avg{n}\approx\frac{1}{3}$, we see a significant reduction 
of density of states compared to those of the flat bands,
which still has a finite density of state attributed to the filling 
of the dispersive band.

Let us look at the quasi-plateau at $\avg{n}\approx\frac{1}{3}$.
Although these newly occupied plaquettes can be arranged to avoid each other as
we did before, they unavoidably will touch the preoccupied ones.
As a result, for each occupied plaquette state, three of its six 
neighbours are occupied alternatively.
The orbital configuration in such a state is as depicted in 
Fig. \ref{fig:dimer}, for each bond shared by two occupied plaquettes, 
the $p$-orbital orientation is parallel to the bond direction as a 
compromise between two neighbouring plaquettes.
The bonding strength exhibits a dimerized pattern.
The ratio between the weakened and strengthened bonds is approximately
$0.44$.
Compared to the gapped dimerized phase discussed below in Sect.
\ref{subsect:strong}, this is a relatively weakly dimerized
phase.

At $\avg{n}>1/2$, all the flat band states are completely filled.
In the weak coupling regime, interaction effects are no longer 
important and crystalline orders vanish.
Near $\avg {n}=1$, the density of states becomes linear with energy
as controlled by the Dirac cones.
Recently, it has been proposed to use the $s$-band in the 
honeycomb optical lattice to simulate the Dirac cone physics \cite{zhu2007}.
The $p_{x,y}$-band Dirac cones described above are also good for this
purpose  and  have even more advantages.
The velocity of the $p_{x,y}$-Dirac cone is much larger than
that of the $s$-band due to a much larger band width 
as shown in Fig. \ref{fig:bandwidth}.
The large energy scale here renders quantum degeneracy 
and the low  temperature regime  much more accessible.

\subsection{Strong coupling regime at $n>\frac{1}{6}$ }
\label{subsect:strong}

\begin{figure}
\centering\epsfig{file=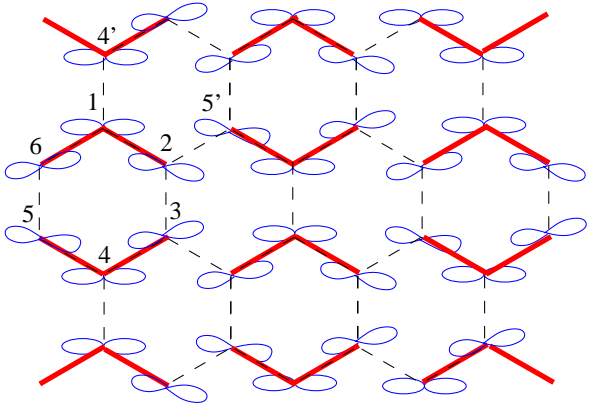,clip=1,width=0.9\linewidth,angle=0}
\caption{The trimerized states at fillings
  $\avg{n}=\frac{1}{3}, \frac{2}{3}$ in the strong coupling regime
  as described by thickened bonds.
  Each trimer contains one particle at $\avg{n}=\frac{1}{3}$
  and two particles at $\avg{n}=\frac{2}{3}$.
}\label{fig:trimer}
\end{figure}

\begin{figure}
\centering\epsfig{file=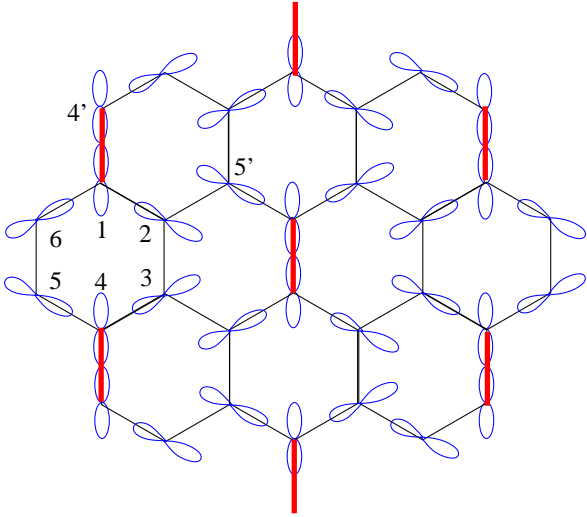,clip=1,width=0.9\linewidth,angle=0}
\caption{The orbital configuration at the filling  $\avg{n}=5/6$ 
exhibits the dimerized state of holes in the strong coupling regime.
Holes mainly distribute on the position of sites $1$ and $4$
in each unit cell with a large bonding strength.
}\label{fig:fvsxth}
\end{figure}

The physics in the strong coupling regime is very different
from that in the weak coupling regime.
Much more crystalline ordered states appear in the strong coupling
regime at commensurate fillings exhibiting rich structures
of dimerization and trimerization orders.
The relation of the filling $\avg{n}$ vs. $\mu$ at $U/t=10$ is depicted
in Fig. \ref{fig:filling} B.
A series of plateaus occur at commensurate fillings of 
$\avg{n}=\frac{i}{6}~(i=1\sim 6)$, which correspond to a set of
charge and bond ordered insulating states.
The charge gap for each insulating state is at the order of $t_\pp$ 
except for that at $\avg{n}=1$ which is at the order of $U$.
Since these gapped state appears at strong interaction regime,
they are not sensitive to a small $t_\perp$.
The band structure described in previous sections is completely 
changed by the strong interactions.
Roughly speaking, at $\avg{n}>\frac{1}{6}$, the preoccupied plaquette 
states exert strong effects to the extra particles and vice versa.
The remaining part of the flat band disappear and the Dirac cone
structure is also destroyed.

At $\avg{n}=\frac{1}{3}$, the strong coupling crystalline ordering
pattern is different from that in the weak coupling regime
depicted in Fig. \ref{fig:dimer}.
The system exhibits trimerized pattern as depicted in Fig. \ref{fig:trimer}.
Each trimer is repented as two thickened bonds and contains one particle.
In other words, each hexagon plaquette in the $\avg{n}=\frac{1}{6}$ case
are occupied by two particles.
Such a state is also three-fold degenerate, and the other two
equivalent states can be obtained by translating the system
one lattice constant to right and left directions.
Let us consider one plaquette unit cell and describe the orbital 
configuration.
We mark the six sites as $1\sim 6$.
The $p$-orbital configurations at sites 1, 6 and 2 are
$p_x$, $\cos\theta p_x \pm \sin \theta p_y (\theta=158.4^\circ)$, 
respectively, and those at 4,  5 and 3 are related
by a reflection operation respect to $x$-axis.
In other words, the occupied $p$-orbital at site 1 is parallel 
to $x$-axis, and that at site 6 is almost along the direction of 
bond $(1, 6)$ with a small deviation of $8.4^\circ$.
The particle density at each site is $n_{6,2,3,5}=0.27$
and $n_{1,4}=0.46$.
The bonding strength between neighouring sites $i,j$ are defined as
$B_{ij}=-\avg{ (p^\dagger_i \cdot \hat e_{ij})(p_j \cdot 
\hat e_{ij})+h.c.}$ . 
There are four non-equivalent bond strengths 
of $(i,j)=(2,1); (2,3); (2, 5^\prime); (1, 4^\prime)$,
where $4^\prime$ and $5^\prime$ are the equivalent sites of 
$4$ and $5$ in the neighbouring plaquettes.
We have $B_{2,1}=0.58 t_\pp$, $B_{2,3}=0.04 t_\pp$, 
$B_{2,5^\prime}=0.14 t_\pp$ and $B_{1,4^\prime}=0$.
The average bonding energy per site can be evaluated from
the above bonding strengths as $0.446t_\pp$.
Instead of the above trimer pattern, one might also think of the dimer 
covering with filling $\frac{1}{3}$ in which only two third of sites 
are covered by dimers.
However, a rough estimation of the average bonding energy per
site is approximately about $\frac{1}{3}t_\pp$, which is 
less energetically favorable because
particles are more localized in the dimer configuration.

The crystalline order pattern at the filling of $\avg{n}=\frac{2}{3}$ 
is similar to that at $\avg{n}=\frac{1}{3}$ with
each trimer containing two particles.
In this case, the parameters above change to $\theta=153^\circ$;
$n_{6,2,3,5}=0.76$, $n_{1,4}=0.49$; 
$B_{2,1}=0.70 t_\pp$, $B_{2,3}=0.07 t_\pp$, $B_{2,5^\prime}=0.12 t_\pp$
and $B_{1,4^\prime}=0.02 t_\pp$.

At $\avg{n}=\frac{1}{2}$, the system exhibits a dimerized
pattern similar to that of $\avg{n}=\frac{1}{3}$
in the weak coupling regime as illustrated in Fig. \ref{fig:dimer}.
The major difference is that the dimerized state here is an 
imcompressible insulating state while that in the weak coupling
regime is with a small but still non-vanishing compressibility.
The dimer is represented by a thickened bond in which one
particle hops back and forth.
It can be considered as a superposition of the two states of two sites
where one is occupied and the other is empty.
There are only two non-equivalent bonding strength:
$B_{1,6}=0.95 t_\pp$  and $B_{1,2}=0.1 t_\pp$.
The former is about one order larger that the latter, thus the system 
is in the strong dimerization limit.
As shown in Fig. \ref{fig:filling} B, the energy scale of this dimerized
phase is set by $t_\pp$, which is much larger than the usual one in
dimerized magnetic systems with $t^2_\pp/U$.

The low energy physics in the dimer phase should be described 
by a quantum dimer model \cite{rokhsar1988}, which includes the 
quantum resonance of different patterns of dimer coverings. 
Although in the physical parameter regime the dimer crystal 
configuration in Fig. \ref{fig:dimer} is stabilized, it would
be interesting to further investigate how to enhance quantum
fluctuations to achieve the quantum disordered dimer liquid phase.
The corresponding possible orbital liquid state in the 
$p_{x,y}$-orbital systems
would be an exciting state for a future study \cite{khaliullin2005}.

The ordering pattern at another commensurate
filling of  $\avg{n}=\frac{5}{6}$ as shown in Fig. \ref{fig:fvsxth}.
The $p$-orbital configurations at sites 1, 6 and 2 are
$p_y$, $\cos\theta p_x \pm \sin \theta p_y $ respectively with
$\theta=150.2^\circ$, and those at 4,  5 and 3 are related
by a reflection operation respect to $x$-axis.
The particle density at each site is $n_{6,2,3,5}=0.94$
and $n_{1,4}=0.62$.
The four non-equivalent bond strengths read
$B_{2,1}=0.36$, $B_{2,3}=0.08$, $B_{2,5^\prime}=0.08$
and $B_{1,4^\prime}=0.83$.
The $\frac{5}{6}$-filling state can be considered as doping 
the insulating state of one particle per site with $\frac{1}{6}$ holes.
Holes are mainly concentrated on the positions of sites 1 and 4 
in each unit cell. 
The corresponding bonds have the largest bonding strength.
Such a state is the dimerized state of holes.

\section{Time of flight spectra}
\label{sect:exp}

Noise correlation has become an important method to detect the ordering
in cold atom systems in optical lattices \cite{altman2004,imambekov2007}.
In all the Mott-insulating states at commensurate fillings described
in figures of \ref{fig:closepack},   \ref{fig:dimer}, \ref{fig:trimer},
and \ref{fig:fvsxth}, the enlarged unit cell contains six sites
forming a plaquette.
They should exhibit themselves in the noise correlation of
the time of flight (TOF) signals.
In the presence of the charge and bond orders, the reciprocal wavevector 
of the reduced Brillouin zone  becomes
$\vec G^\prime_1=(\frac{4\pi}{3\sqrt 3 a},0)=\frac{-1}{3}\vec G_1
+\frac{2}{3} \vec G_2 $ 
and $\vec G_2^\prime=(\frac{-2\pi}{3\sqrt 3 a}, \frac{2\pi}{3a})
=\frac{2}{3} \vec G_1-\frac{1}{3} \vec G_2$, where
$\vec G_{1,2}$ are the reciprocal wavevectors for the original Brillouin
zone.
The correlation function is defined as
\bea
C_t(\vec r, \vec r^\prime)=\avg{n(\vec r) n(\vec r^\prime)}_t 
-\avg{n(\vec r)}_t \avg{n(\vec r^\prime)}_t,
\label{eq:noise}
\eea
where $t$ is the flying time.

For the close pack hexagon state at $\avg{n}=\frac{1}{6}$, Eq. \ref{eq:noise}
can be easily calculated.
We have
$\avg{n(\vec r)_t}=(\frac{m}{\hbar t})^3 |\psi(\vec k)|^2$,
where $\vec k=m \vec r/(\hbar t)$, and $\psi(\vec k)$
is the Fourier transform of the plaquette-Wannier state
depicted in Fig. \ref{fig:wannier}.
Thus 
\bea
C_t(\vec r, \vec r^\prime)&=& \mp\frac{N}{6}  (\frac{m}{\hbar t})^6
|\psi(\vec k)|^2 |\psi(\vec k^\prime )|^2  \nn \\
&\times& \sum_{\vec G^\prime}
\delta(\vec k-\vec k^\prime -\vec G^\prime),
\eea
where  
$'-'$ ($'+'$) is for fermions (bosons)  respectively, 
$\vec G^\prime = m \vec G^\prime_1 
+ n \vec G^\prime_2$ with $m,n$ integers, and
$\vec k^\prime=m \vec r^\prime/(\hbar t)$.
After a spatial averaging and normalization, we find
\bea
C_t(\vec d)&=&\int d \vec r \frac{C_t(\vec r +\frac{\vec d}{2},
\vec r -\frac{\vec d}{2})}
{\avg{n(\vec r+ \frac{\vec d}{2})}_t \avg{n(\vec r
-\frac{\vec d}{2})}_t}\nn \\
&\propto &\mp\sum_{\vec G^\prime}\delta (\vec k -\vec G^\prime),
\label{eq:timeofflight}
\eea
where $\vec k=m \vec d/(\hbar t)$.
All the $\delta$-peaks are with equal weight because of the cancellation
of the Fourier transform of the Wannier function, and the six-fold
rotational symmetry in Fig. \ref{fig:wannier}.

For the crystalline ordering of fermions at other commensurate fillings,
the noise correlation functions still exhibit the $\delta$-peaks
located at the same reciprocal wavevectors of the reduced Brillouin zone.
However the form factors are more complicated.
Generally, $\avg{n(\vec r)}$ and $C_t(\vec r, \vec r^\prime)$
can be calculated as
\bea
\avg{n(\vec r)_t}&\propto&\sum_{\mu\nu}
\psi^*_\mu(\vec k) \psi_\nu(\vec k)\avg{\Omega|
p^\dagger_\mu(\vec k) p_\nu(\vec k)|\Omega}, \nn \\
C_t(\vec r, \vec r^\prime)&\propto&  - |\sum_{\mu\nu}
\psi^*_\mu(\vec k)\psi_{\nu}(\vec k^\prime)
\avg{\Omega|p^\dagger_\mu(\vec k) p_{\nu}(\vec k^\prime)|\Omega}|^2
\nn \\
&\times& \sum_{\vec G^\prime}
\delta(\vec k-\vec k^\prime -\vec G^\prime),
\eea
where the Greek indices $\mu$ and $\nu$
denote the Wannier functions for the 12 $p_{x,y}$ orbitals in
one plaquette.
In particular, due to the loss of the six-fold lattice rotational
symmetry in \ref{fig:trimer} and \ref{fig:fvsxth}, the noise
spectra of $C_t(\vec d)$ should show the reduced two fold rotational symmetry.
Fig. \ref{fig:dimer} still keeps the six-fold rotational symmetry,
but the weight of the $\delta$-functions should not be the same
as in Eq. \ref{eq:timeofflight}.

\section{Conclusion and Discussion}
\label{sect:conclusion}
In summary, we have proposed the laboratory analog simulation of a
new kind of artificial graphene, unavailable in nature, where the
$p_{x,y}$-orbitals are the key, unlike the real graphene made of 
the $p_z$-orbital.
This switching of orbitals, as shown in this work, lead to novel 
strong correlation physics which cannot be studied in the corresponding 
solid state graphene systems.

We have shown the band structure of $p_{x,y}$-orbital honeycomb lattices
contains both Dirac cones and flat bands.
Particle interactions stabilize various incompressible
Wigner crystal-like states at commensurate fillings.
In particular, we have described the exact many body ground 
state at $\avg{n} =\frac{1}{6}$, which exhibits close
packed hexagon plaquette order.
Various charge and bond orderings appear in the strong
coupling regime at higher commensurate fillings.
These states exhibit their patterns in the noise correlation 
of time of flight experiments.
Taking into account the recent exciting experimental realization
of the $p$-orbital bosons \cite{mueller2007} and the fact
that the honeycomb optical lattices were experimentally constructed
quite some time ago \cite{grynberg1993},
the $p_{x,y}$-orbital counterpart of graphene may be
achieved in the laboratory in the near future.

Let us compare the Wigner crystal states in the $p_{x,y}$-orbital systems
with those in the electron gas systems.
Quantum Monte-Carlo simulations show that the Wigner crystal state
is stable in the very low density regime at $r_s>39$ in two 
dimensions, where $r_s$ is the ratio between the average inter-electron 
distance and the Bohr radius \cite{tanatar1989}.
The long range Coulomb interactions dominate over 
the kinetic energy when $r_s$ is large.
In contrast, even the shortest range repulsive interaction can stabilize 
the crystal state in the $p_{x,y}$-orbital honeycomb lattice
due to the suppression of the kinetic energy by the band flatness.
Wigner crystal state also occurs in the fractional quantum Hall systems
due to the suppression of kinetic energy by the magnetic field
\cite{chen2006,ye2002}.
At low filling factors, crystalline ordered states energetically
win over the Laughlin liquid state.
It is also interesting to note the difference between our system
and the $p_z$-orbital system of graphene, where
the characteristic ratio between Coulomb interaction
and kinetic energy $\frac{e^2}{\hbar v_f}$ ($v_f$ is the slope  of the 
Dirac cone) is a constant independent of charge carrier density.
As pointed out in Ref. \cite{dahal2006,dassarma2007,tse2007}, interactions in
graphene are not strong enough to stabilize the Wigner crystal state 
at any density.

Many interesting problems still remain open for further exploration,
and we will leave them in future publications.
For example, for the spinful fermions with repulsive interactions,
it is natural to expect ferromagnetism due to the flat band structure.
It would be interesting to study the competition between antiferromagnetic
exchange and flat band ferromagnetism.
If interactions are attractive, the pairing problem and
the corresponding BCS-BEC crossover  in the flat band
might prove interesting.
On the other hand, if we load bosons into the flat band beyond the
density of $\avg{n}=\frac{1}{6}$, the frustration effect due to
the band flatness to the superfludity is a challenging problem.
Most intriguing is the possibility of exotic incompressible 
states analogous to the Laughlin
liquid in fractional quantum Hall effect. These cannot be captured
within the mean-field approximation used here for $n>1/6$. If one could
devise appropriate variational liquid states projected into the flat
band, these could be compared energetically with the Wigner crystals
found here. Given the richness and surprises encountered in the 
fractional quantum Hall effect,
flat band physics in optical lattices appears rife with possibility.

\begin{acknowledgements}
C. W. thanks L. M. Duan, E. Fradkin, and T. L. Ho
for helpful discussions, and especially L. Balents
and D. Bergman for an early collaboration.
C. W. is supported by the start up funding at University of California,
San Diego and the Sloan Research Fellowship.
S. D. S. is supported by  ARO-DARPA. 
\end{acknowledgements}


\end{document}